\def\etal{{et~al.}}
\documentclass[10pt,preprint]{aastex}
\usepackage{subfigure}
\begin{document}
\setcounter{figure}{0}
\title{Proper Motions of Dwarf Spheroidal
Galaxies from \textit{Hubble Space Telescope} Imaging.  IV:
Measurement for Sculptor.\footnote{Based on observations with NASA/ESA
\textit{Hubble Space Telescope}, obtained at the Space Telescope
Science Institute, which is operated by the Association of
Universities for Research in Astronomy, Inc., under NASA contract NAS
5-26555.}}

\author{Slawomir Piatek} \affil{Dept. of Physics, New Jersey Institute
of Technology,
Newark, NJ 07102 \\ E-mail address: piatek@physics.rutgers.edu}

\author{Carlton Pryor}
\affil{Dept. of Physics and Astronomy, Rutgers, the State University
of New Jersey, 136~Frelinghuysen Rd., Piscataway, NJ 08854--8019 \\
E-mail address: pryor@physics.rutgers.edu}

\author{Paul Bristow}
\affil{Space Telescope European Co-ordinating Facility, 
Karl-Schwarzschild-Str. 2, D-85748, Garching bei Munchen, Germany \\
E-mail address: bristowp@eso.org }

\author{Edward W.\ Olszewski}
\affil{Steward Observatory, The University of Arizona,
    Tucson, AZ 85721 \\ E-mail address: eolszewski@as.arizona.edu}

\author{Hugh C.\ Harris}
\affil{US Naval Observatory, Flagstaff Station, P. O. Box 1149,
Flagstaff, AZ 86002-1149 \\ E-mail address: hch@nofs.navy.mil}

\author{Mario Mateo} \affil{Dept. of Astronomy, University of
Michigan, 830 Denninson Building, Ann Arbor, MI 48109-1090 \\
E-mail address: mateo@astro.lsa.umich.edu}

\author{Dante Minniti}
\affil{Universidad Catolica de Chile, Department of Astronomy and
Astrophysics, Casilla 306, Santiago 22, Chile \\
E-mail address: dante@astro.puc.cl}

\author{Christopher G.\ Tinney}
\affil{Anglo-Australian Observatory, PO Box 296, Epping, 1710,
Australia \\ E-mail address: cgt@aaoepp.aao.gov.au}

\begin{abstract}

	This article presents a measurement of the proper motion of
the Sculptor dwarf spheroidal galaxy determined from images taken with
the Hubble Space Telescope using the Space Telescope Imaging
Spectrograph in the imaging mode.  Each of two distinct fields
contains a quasi-stellar object which serves as the ``reference
point.''  The measured proper motion of Sculptor, expressed in the
equatorial coordinate system, is ($\mu_{\alpha}, \mu_{\delta})=(9 \pm
13,2 \pm 13)$~~mas~century$^{-1}$.  Removing the contributions from
the motion of the Sun and the motion of the Local Standard of Rest
produces the proper motion in the Galactic-rest-frame:
$(\mu_{\alpha}^{\mbox{\tiny{Grf}}},\mu_{\delta}^{\mbox{\tiny{Grf}}}) =
(-23 \pm 13, 45 \pm 13)$~~mas~century$^{-1}$.  The implied space
velocity with respect to the Galactic center has a radial component of
$V_{r}=79
\pm 6$~km~s$^{-1}$ and a tangential component of $V_{t}=198 \pm
50$~km~s$^{-1}$.  Integrating the motion of Sculptor in a realistic
potential for the Milky Way produces orbital elements.  The
perigalacticon and apogalacticon are 68~$(31,83)$~kpc and
122~$(97,313)$~kpc, respectively, where the values in the parentheses
represent the $95\%$ confidence interval derived from Monte Carlo
experiments.  The eccentricity of the orbit is $0.29~(0.26,0.60)$ and
the orbital period is $2.2~(1.5,4.9)$~Gyr.  Sculptor is on a polar
orbit around the Milky Way: the angle of inclination is
$86~(83,90)$~degrees.

\end{abstract}


\section{Introduction}
\label{sec:intro}

	Shapley (1938) discovered the Sculptor dwarf spheroidal (dSph)
galaxy --- the first example of this type of galaxy in the vicinity of
the Milky Way --- on a plate with a 3~hour exposure time taken with the
Bruce telescope.  Shapely notes ``... that systems such as the Sculptor
cluster may not be uncommon; their luminosity characteristics would
enable them to escape easy discovery.''  Since the detection of
Sculptor, astronomers have identified eight other dSphs.

	Sculptor is at a celestial location of $(\alpha,\delta) =
(01^{\mbox{h}}00^{\mbox{m}}09^{\mbox{s}}, -33^{\circ}
42^{\prime}30^{\prime\prime})$ (J2000.0; Mateo 1998), which corresponds
to Galactic coordinates of $(\ell, b)=(287\fdg 5,-83\fdg 2)$.  Thus,
Sculptor lies nearly at the South Galactic Pole.

	Kaluzny \etal\ (1995) searched for variable stars in a
$15^{\prime} \times 15^{\prime}$ field centered approximately on the
dSph by taking $V-$ and $I-$band images with the 1-m Swope telescope
at Las Campanas Observatory over a period of more than two months.
The search resulted in the identification of 226 RR~Lyr stars.  The
average $V-band$ magnitude of the RR~Lyr stars gives a distance
modulus of $(m-M)_{V}=19.71$, which corresponds to a heliocentric
distance of 87~kpc.  This estimate is practically the same as that
obtained by Hodge (1965); it is consistent with the estimate of Baade
\& Hubble (1939), but somewhat larger than the estimate of Kunkel
\& Demers (1977).  This study adopts the estimate of Kaluzny \etal\
(1995) for the distance to Sculptor.
	
	Irwin \& Hatzidimitriou (1995) derives the most comprehensive
set of structural parameters for Sculptor --- and seven other dSphs ---
using star counts from UK Schmidt telescope plates.  With a luminosity
of $(1.4\pm 0.6) \times 10^{6}$~L$_{\odot}$, Sculptor is among the most
luminous dSphs.  Its major-axis core and limiting radii are $5.8 \pm
1.6$~arcmin and $76.5 \pm 5.0$~arcmin, respectively, which are in good
agreement with the values derived by Demers, Kunkel, \& Krautter
(1980).  However, they differ from the values derived by Eskridge
(1988a), who also uses star counts from UK Schmidt telescope plates.
The discrepancy is likely due to an underestimation of the central
density in the latter study for reasons that are discussed in Irwin \&
Hatzidimitriou (1995).  The isopleth map of Sculptor (Panel (f) of
Figure~1 in Irwin \& Hatzidimitriou 1995) shows that the ellipticity of
the isodensity contours increases with increasing projected radius from
the center of the dSph: the ellipticity is consistent with 0 in the
inner 10~arcmin and smoothly increases to a value of 0.32 in the
outermost region.  The study observes that Sculptor ``... looks
remarkably similar to numerical simulations of dSph galaxies that are
tidally distorted.'' The position angle of the major axis is $99 \pm
1$~degrees.  Eskridge (1988b) finds asymmetric ``structure'' in
Sculptor in an isopleth map of the difference between the stellar
surface density and a fitted 2-D model.  In contrast, Irwin \&
Hatzidimitriou (1995) finds only the increase of the ellipticity with
projected radius in a similar map.  The left panel in
Figure~\ref{fig:fields} shows a 30~arcmin $\times$ 30~arcmin region of
the sky centered on Sculptor.  The dashed ellipse is the boundary of
the core.

	Walcher \etal\ (2003) studies the structure of Sculptor ---
together with those of Carina and Fornax --- using $V$-band images
taken with the MPG/ESO 2.2-m telescope at La Silla.  The images have an
areal coverage of 16.25~square degrees and reach a limiting magnitude of $V
\approx 23.5$.  The study derives major-axis core and limiting radii of
$7.56 \pm 0.7$~arcmin and $40 \pm 4$~arcmin, respectively, using a
King (1962) model, as did Irwin \& Hatzidimitriou (1995).  The
1-$\sigma$ disagreement between the two derived core radii is perhaps
larger than expected from two data sets that have many stars in
common.  The apparently more serious disagreement between the two
limiting radii is most likely due only to a larger true uncertainty in
the limiting radius caused by uncertainties in the background surface
density and the poor fit of the model to the outer part of the surface
density profile.  Walcher \etal\ (2003) confirms that the ellipticity
of the surface-density contours increases with increasing projected
radius; the contours also suggest ``extensions'' from the ends of the
major axis that are interpreted as tidal tails.  The radial projected
density profile shows a ``break'' --- a departure from the fitted King
model --- at around 30~arcmin, which the study interprets as evidence
for the existence of an extended stellar component.  Walcher \etal\
(2003) uses the relation between the King-model tidal radius, the
mass, and the perigalacticon of a Galactic satellite developed by Oh,
Lin, \& Aarseth (1992) to deduce that Sculptor has a perigalacticon of
28~kpc.

	A recent article by Coleman \etal\ (2005) does not confirm the
finding in Walcher \etal\ (2003) that Sculptor has tidal tails and an
extended stellar component.  Instead, the analysis of photometric data
in the $V$ and $I$ bands for a $3.1^{\circ}\times3.1^{\circ}$ field
shows that a King model with a limiting radius of $72.5\pm4.0$~arcmin
is a satisfactory fit to the radial profile of stars that lie on the
giant branch.  The limiting magnitudes of the photometry are $V=20$ and
$I=19$, respectively.  The study notes that oversubtracting the field
population in its data produces a radial profile with the smaller
fitted limiting radius and significant extratidal structure found in
Walcher \etal\ (2003).  Using additional information from the
spectroscopy of 723 stars selected from the red giant branch, Coleman
\etal\ (2005) derives an upper limit of $2.3\% \pm 0.6\%$ for the
contribution from stars beyond the tidal boundary to the total mass of
Sculptor.  The study does not find any conclusive evidence for tidal
interaction between the Milky Way and Sculptor.

	In contrast, Westfall \etal\ (2005) does find evidence.  This
study uses imaging in the $M$, $T_{2}$, and $DDO51$ bands to separate
member giants from foreground dwarfs in a 7.82 degrees$^{2}$ area that
covers the eastward side of Sculptor including the central region.
Candidate members are also selected from the region of the blue
horizontal branch in the color-magnitude diagram.  The selection of
members is checked with spectroscopy for 147 candidates.  The study
finds members up to 150 arcmin from the center of Sculptor --- the
spatial extent of the survey and beyond the tidal boundary if that is
identified with the measured King limiting radius of 80~arcmin.
Several of these stars are spectroscopically-confirmed members.  The
radial surface brightness profile shows a break to a shallower slope
at a radius of about 60~arcmin, which resembles the radial profiles
seen in simulations of satellites interacting with the Milky Way
(Johnston \etal\ 1999).  Thus, Westfall \etal\ (2005) argues in favor
of a significant tidal interaction between the Milky Way and Sculptor.
It is beyond the scope of this work to resolve the apparent conflict
between Coleman \etal\ (2005) and Westfall \etal\ (2005) by judging
the merits of the analyses presented in both articles.  Needless to
say, a disagreement exists about the effect of the Galactic tidal
field on the structure of Sculptor; measuring the proper motion of the
dSph may allow us to impose constraints on this effect.

	Armandroff \& Da Costa (1986) measured the radial velocities
of 16 giants in Sculptor which average to produce a systemic
heliocentric velocity of $107.4 \pm 2.0$~km~s$^{-1}$.  This
measurement alleviated the large uncertainty in this quantity, which
existed due to mutually contradictory estimates from Hartwick \&
Sargent (1978) and Richter \& Westerlund (1983).  More recently,
Queloz, Dubath, \& Pasquini (1995) measured the radial velocities of
23 giant stars.  The sample includes 15 stars observed previously by
Armandroff \& Da Costa (1986).  The implied systemic heliocentric
velocity of $109.9 \pm 1.4$~km~s$^{-1}$ (after excluding two stars
that are likely binaries) agrees within the quoted uncertainties with
the measurement of Armandroff \& Da Costa (1986).  Our article adopts
a mean velocity of $109.9 \pm 1.4$~km~s$^{-1}$ for calculating the
space velocity of Sculptor.  Queloz, Dubath, \& Pasquini (1995) finds
no apparent rotation of the dSph around its minor axis.  Tolstoy
\etal\ (2004) measured radial velocities for 308 potential members of
the dSph and find a systemic velocity of 110~km~s$^{-1}$.  There is no
discussion of rotation, but Figure~4 in that article shows that the
velocity dispersion does not increase with radius, as would be
expected if there were a net rotation larger than the central
dispersion.  Interestingly, the study finds that the red horizontal
branch stars have a more compact spatial distribution and a smaller
velocity dispersion than the older and more metal-poor blue horizontal
branch stars.  The two most recent photometric and spectroscopic
surveys by Coleman \etal\ (2005) and Westfall \etal\ (2005) confirm the
greater central concentration of the more metal-rich stars.  The two
studies also find no evidence for rotation.

	The mass-to-light ratio, $(M/L)$, of Sculptor is larger than a
typical value for a Galactic globular cluster; it is, however, smaller
than the $M/L$s for some other Galactic dSphs.  Armandroff \& Da Costa
(1986) derives a central $M/L_{V}$ of $6.0 \pm 3.1$ and Queloz, Dubath,
\& Pasquini (1995) determines the somewhat larger value of $13 \pm 6$,
in solar units.  Both studies note that the measured $M/L$ does not
imply unequivocal support for dark matter in Sculptor.

	The stars of Sculptor are old.  Fitting isochrones to the
principal sequences in the color-magnitude diagram, Da Costa (1984)
finds that the majority of the stars are younger by ``2--3~Gyr than
Galactic globular clusters of similar metal abundance provided the
helium abundances and the CNO/Fe ratios are also similar.''  Da Costa
(1984) also detects ``blue stragglers'' and estimates their age to be
about 5~Gyr under the assumption that they are ``normal'' main-sequence
stars, i.e., stars which did not acquire mass from a companion.  No
stars younger than 5~Gyr exist in Sculptor, indicating an absence of
ongoing or recent star formation.  However, Sculptor contains HI gas.

	Carignan \etal\ (1998) and later Bouchard \etal\ (2003) detect
two distinct clouds of HI that are diametrically opposite to each other
almost along the minor axis and 20 -- 30~arcmin from the center of the
dSph.  These clouds are within the tidal radius.  The HI gas is very
likely to be associated with Sculptor because its mean heliocentric
velocity is similar to that of the dSph.

	Carignan \etal\ (1998) discusses mechanisms that might account
for the existence of the clouds.  Removing the gas from the dSph by a
time-dependent tidal force due the Milky Way is one possibility.
Carignan \etal\ (1998) suggests that the alignment between the proper
motion vector from Schweitzer \etal\ (1995) and a line passing through
the two clouds supports this hypothesis for the origin of the clouds.
However, if the tidal force affects the HI, it should also affect the
stars and the possible signatures of tides in the stellar component of
Sculptor are either absent or inconsistent with the direction of the
Schweitzer \etal\ (1995) proper motion vector.  For example, the
increasing ellipticity of isodensity contours with increasing
projected radius could be due to tides (e.g., Johnston, Spergel, \&
Hernquist 1995), but then the major axis should be along the proper
motion vector.  The possible ``tidal extensions'' reported by Walcher
\etal\ (2003) are at the ends of the major axis.  If these extensions
are a continuation of the increasing ellipticity, they also argue for
an orbital plane parallel to the major axis.  Walcher \etal\ (2003)
claims that the eastern extension bends to the south, i.e., parallel to
the minor axis and so argues that the orbital plane is aligned in the
north-south direction.  However, this alignment is inconsistent with
the increasing ellipticity being due to the tidal force since, given
the large distance of Sculptor, the Sun is nearly in the orbital plane
and so the increasing ellipticity should be aligned with the tidal
extensions.

	Schweitzer \etal\ (1995) reports the first measurement of the
proper motion for Sculptor: $(\mu_{\alpha},\mu_{\delta})=(72 \pm 22,
-6 \pm 25)$~mas~century$^{-1}$.  This value includes contributions
from the motions of the Sun and LSR; this article refers to this
quantity as the ``measured proper motion."  The measurement derives
from 26 photographic plates imaged with a variety of ground-based
telescopes using either a ``blue'', B, or V filter.  The earliest
epoch is 1938 and the latest is 1991.  The study estimates, among
other quantities, the perigalacticon of the implied orbit.  The best
estimate ranges from 60~kpc for the ``infinite halo'' potential of the
Milky Way to 78~kpc for the ``point mass'' potential.  If the
perigalacticon is no smaller than 60~kpc, then the Galactic tidal
force has not played a significant role in the evolution of Sculptor.
Numerical simulations of Piatek \& Pryor (1995) or Oh, Lin, \& Aarseth
(1995) show that for a typical dSph, even with a $M/L_V$ as low as 3,
a perigalacticon of 60~kpc is too large for tides to have an important
effect.

	Motivated by the idea that some of the Galactic dSphs and
globular clusters may be pieces of a tidally-disrupted progenitor
satellite galaxy, several studies propose that they form ``streams''
in the Galactic halo.  Lynden-Bell (1982) hypothesizes that Fornax,
Leo I, Leo II, and Sculptor are members of the ``FLS stream.''
Majewski (1994) adds the newly-discovered Sextans to the FLS stream
and recalculates its common plane --- naming it the ``FL$^{2}$S$^{2}$
plane.''  The FLS and the FL$^{2}$S$^{2}$ planes differ only slightly.
In a more extensive study, Lynden-Bell \& Lynden-Bell (1995) infers
that Sculptor may belong to one of three possible streams (see their
Table~2).  Stream \# 2 contains the LMC, SMC, Draco, Ursa Minor,
and, possibly, Sculptor and Carina; stream \# 4a contains Sextans,
Sculptor, Pal 3, and, possibly, Fornax; finally, stream \# 4b
contains Sextans, Sculptor, and, possibly, Fornax.  For each stream,
Lynden-Bell \& Lynden-Bell (1995) calculates the expected proper
motion of Sculptor.

	Kroupa \etal\ (2004) notes that the 11 dwarf galaxies nearest
to the Milky Way form a disk with a thickness to radius ratio of
$\leq$~0.15.  The article argues that the distribution expected for
such nearby substructure in a cold-dark-matter universe is spherical,
that the observed distribution is not, and, thus, that these objects
are the tidal debris from the disruption of a larger satellite galaxy.
In contrast, Kang \etal\ (2005) and Zentner \etal\ (2005) find that a
planar distribution of nearby Galactic satellites is actually common
in numerical simulations of galaxy formation.  A direct comparison
between the results of the simulations and the distribution of nearby
satellites finds that they are consistent.

	A test of the reality of streams or planar alignments is to
measure the space motions of the satellites.  Piatek \etal\ (2005;
P05) reports a proper motion for Ursa Minor.  The implied orbit for
Ursa Minor is not in the plane defined by Kroupa \etal\ (2004).  The
proper motion also rules out membership in the stream proposed by
Lynden-Bell \& Lynden-Bell (1995).  The measured proper motion for
Carina (Piatek \etal\ 2003; P03) does not agree well with the
predictions of Lynden-Bell \& Lynden-Bell (1995), but is not precise
enough to rule out membership in a stream.  Piatek \etal\ (2002; P02)
finds that a preliminary proper motion for Fornax is inconsistent with
the predictions of Lynden-Bell \& Lynden-Bell (1995) and that its
direction is also inconsistent with an orbit in the FL$^{2}$S$^{2}$
plane.  Dinescu \etal\ (2004) reports an independent measurement of
the proper motion of Fornax.  This motion is consistent, within its
uncertainty, with the predictions of Lynden-Bell \& Lynden-Bell (1995)
and the direction of this motion is along the great circle defined by
the FL$^{2}$S$^{2}$ plane.  Dinescu \etal\ (2004) notes that the
proper motion for Sculptor in Schweitzer \etal\ (1995) is inconsistent
with the FL$^{2}$S$^{2}$ plane.

	This article reports a second independent measurement of the
proper motion for Sculptor and discusses the implications of the
derived space motion on the dSph-Galaxy interaction.
Section~\ref{sec:data} describes observations and the data.  The
following section describes the analysis of the data leading to the
derivation of the proper motion.  Section~\ref{sec:pmm} compares the
proper motion from Schweitzer \etal\ (1995) with the one reported in
this article.  The next section, Section~\ref{sec:orbit}, integrates
and describes the orbit of Sculptor.  Section~\ref{sec:disc} discusses
the implications of the orbit for the importance of the Galactic tidal
force on the structure and internal kinematics of Sculptor, for the
star formation history, and for the membership of Sculptor in the
proposed streams of galaxies and globular clusters in the Galactic
halo.  The final section is a summary of the main results and
conclusions.

\section{Observations and Data}
\label{sec:data}

	The Hubble Space Telescope (HST, hereafter) imaged two
distinct fields in Sculptor using the Space Telescope Imaging
Spectrograph (STIS, hereafter) in imaging mode with no filter (50CCD).
Each field contains a known quasi-stellar object (QSO, hereafter),
which serves as a reference point.  The left panel of
Figure~\ref{fig:fields} depicts the locations of the two fields on the
sky: two small squares --- one inside and the other outside of the
core.  The name of the field inside the core is SCL~$J0100-3341$,
which derives from the IAU designation of the QSO in this field.
Tinney \etal\ (1997) confirms the identity of this QSO: it is at
$(\alpha, \delta) = (01^{\mbox{h}}00^{\mbox{m}}25\fs 3, -33^{\circ}
41^{\prime}07^{\prime\prime})$ (J2000.0), has a redshift $z=0.602 \pm
0.001$, and has a magnitude $B=20.4$.  The observations of the
SCL~$J0100-3341$ field occurred on September 24, 2000 and on September
26, 2002.  At each epoch, there are three exposures at each of the
eight dither pointings for the total of 24 images.  The
``ORIENTAT'' angle --- the position angle of the Y axis of the CCD
measured eastward from north --- is the same to within one-tenth of a
degree for all of the exposures and equal to -67.5 degrees.  The
top-right panel in Figure~\ref{fig:fields} shows the SCL~$J0100-3341$
field.  The QSO is in the cross-hair.

	The name of the field outside of the core is SCL~$J0100-3338$.
The QSO in this field, also confirmed by Tinney \etal\ (1997), is at
$(\alpha, \delta) = (01^{\mbox{h}}00^{\mbox{m}}32\fs 6, -33^{\circ}
38^{\prime}32^{\prime\prime})$ (J2000.0), has a redshift $z=0.728 \pm
0.001$, and has a magnitude $B=20.4$.  $HST$ observed this field on
September 13, 1999; September 28, 2000; and on September 28, 2002.  At
each of the three epochs, there are three exposures at each of the
eight dither pointings for a total of 24 images.  The ``ORIENTAT''
angle is the same to within one-tenth of a degree and equal to -69.3
degrees for all of the exposures for this field.  The bottom-right
panel of Figure~\ref{fig:fields} shows the SCL~$J0100-3338$ field.
The QSO is in the cross-hair.  Owing to its greater distance from the
center of the dSph, the SCL~$J0100-3338$ field contains fewer stars
than does the SCL~$J0100-3341$ field.

	Bristow (2004) and P05 discuss the effect of the decreasing
charge transfer efficiency of the STIS CCD on astrometric
measurements.  If not accounted for, the decreasing charge transfer
efficiency may introduce a spurious contribution to a measured proper
motion.  Bristow \& Alexov (2002) developed computer software which
approximately restores an image taken with STIS to its pre-readout
condition.  All of the results that this article reports are based on
images restored using the program of Bristow \& Alexov (2002).

\section{Analysis}
\label{sec:analysis}

	P02 describes our method of deriving a proper motion from
images taken with \textit{HST} and containing at least one QSO.
Fundamental to the method is the concept of an effective point-spread
function (ePSF, hereafter), which Anderson \& King (2000) describes in
detail.  The subsequent two articles in this series, P03 and P05,
expand and improve upon the basic method.  The analysis reported here
incorporates only minor new features into the method; thus, the reader
should consult those earlier articles for the details.  Instead, this
study mentions the major elements of the method alongside figures
depicting key diagnostics of the performance of the method and briefly
describes the new features.

\subsection{Flux Residuals}
\label{sec:rf}

	Equation 22 in P02 defines a ``flux residual'' diagnostic,
$\cal{RF}$.  It is the measure of how the shape of the constructed
ePSF matches the shape of an image of an object.  In the case of a
perfect match, ${\cal RF} =0$; if the ePSF is narrower, ${\cal RF} >
0$; otherwise, ${\cal RF} < 0$.

	Several factors affect the shape of the PSF for an object.  1.
Type of an object.  A PSF for a galaxy is generally wider than that
for a star, all else being equal.  2. Color of an object.  Because of
diffraction and aberrations, the width of the PSF is color-dependent.
3. Tilt or curvature of the focal plane.  The PSF varies with location
because the CCD surface and focal plane do not coincide everywhere.
4. Thermal expansion.  Because the \textit{HST} moves in and out of
the Earth's shadow, its temperature is continuously changing.  These
changes cause the telescope to expand or contract, affecting its focal
length.  5. Charge traps in the CCD.  As the packets of charge
representing an object move along the $Y$ axis (the direction of
readout for STIS), those on its leading side fill partially each trap
encountered, so that there are fewer traps available to remove charge
from subsequent packets (Bristow \& Alexov 2002).  This non-uniform
loss of charge across the object changes its PSF.

	Given the aforementioned factors affecting the shape of the
PSF, a plot of ${\cal RF}$ \textit{versus} the $X$- or $Y$-coordinate
of an object will, in the best case, show that the points scatter
around ${\cal RF}=0$.  In a less desirable case, the points may show
trends with $X$ or $Y$ or both.  These trends signal that the true PSF
varies with location.

	Because of the scarcity of stars in the observed fields, our
method constructs a single and constant ePSF for a given field and
epoch. A constant ePSF is one that does not vary with either $X$ or
$Y$.  Figures \ref{fig:rf-scl1} and \ref{fig:rf-scl2} show plots of
${\cal RF}$ \textit{versus} $X$ (panels in the left-hand column) and
${\cal RF}$ \textit{versus} $Y$ (panels in the right-hand column) for
the SCL~$J0100-3341$ and SCL~$J0100-3338$ fields, respectively.  The
rows of panels from top to bottom are each one epoch, arranged in
chronological order.  The filled squares in a plot correspond to the
QSO.  Note that the number of ${\cal RF}$ values for a given object
may be equal to the number of exposures --- individual images --- at a
given epoch, or be less if the object is not measured in one or more
exposures.

	No panel in Figure~\ref{fig:rf-scl1}, except for the top-left
one, shows a trend between ${\cal RF}$ and $X$ or ${\cal RF}$ and $Y$.
The top-left panel shows that the mean ${\cal RF}$ decreases linearly
with $X$, implying that the shape of the true PSF becomes
progressively narrower and more peaked than that of the constructed
ePSF with increasing $X$.  We are unable to trace the origin of this
dependence.  The values of ${\cal RF}$ for the QSO are larger than
those for other objects at both epochs and are all positive, implying
that the true PSF for the QSO is wider than the constructed ePSF and
than that for a star.

	No panel in Figure~\ref{fig:rf-scl2} shows a trend as
conspicuous as the one in the upper-left panel of
Figure~\ref{fig:rf-scl1}.  Nevertheless, the left-hand panel in the
middle row does show a hint of variability of the true PSF with
location.  The PSF of the QSO in the SCL~$J0100-3338$ is similar to
that of a star.  The values of ${\cal RF}$, though still biased towards
positive values, are comparable to those for bright stars.

	There are two reasons why the PSF of a QSO can be different
from that of a star.  1. The underlying galaxy can broaden the image
of a QSO.  2. The color of a typical QSO is bluer than that of a
typical star.  So, particularly for the unfiltered STIS imaging, the
true PSF is narrower for a bluer object.  Thus, depending on the
interplay between the distance to a QSO and its color, the values of
${\cal RF}$ for the QSO can average more positive than, more negative
than, or the same as those for a bright star.  Visual inspection of
the QSO in the SCL~$J0100-3341$ field shows what appears to be a
single spiral arm or tidal feature extending from its image,
suggesting that the underlying galaxy is indeed the cause of the large
positive values of ${\cal RF}$ for this QSO.

	Experience with the data for other dSphs (P02, P03, and P05)
has shown that trends in the ${\cal RF}$ values with position do not
necessarily produce systematic errors in the positions of objects.  The
next section searches for such systematic errors in the position.

\subsection{Position Residuals}
\label{sec:rx-ry}

	Fitting an ePSF to the science data array of an object (the
$5\times5$ array of pixels representing an object; see P02 for more
detail on this array and our procedures) determines its centroid.
With 24 images per field and epoch, there can be up to 24 measurements
of the centroid.  The actual number will be smaller than 24 if an
object is flagged out from one or more images because its array is
corrupted by cosmic rays or hot pixels.  The dithering, rotation, and
change of scale (e.g., due to ``breathing'' of the \textit{HST})
between any two images cause the centroid of an object measured in
these two images to differ.  Therefore, at each epoch, every field has
a fiducial coordinate system that coincides with the coordinate system
of the first image in chronological order.  The adopted transformation
from the coordinate system of each subsequent image to the fiducial
system contains a linear translation, rigid rotation, and a uniform
scale change.  Let $(X_{0,j}^{i,k}, Y_{0,j}^{i,k})$ be the centroid of
object $i$ at epoch $j$ in image $k$ transformed to the fiducial
coordinate system and the mean centroid of object $i$ in the fiducial
coordinate system of epoch $j$ be $(<X_{0,j}>^{i},<Y_{0,j}>^{i})$.
Define position residuals, ${\cal RX}_{j}^{i,k}$ and ${\cal
RY}_{j}^{i,k}$, for an object $i$ as ${\cal
RX}_{j}^{i,k}=<X_{0,j}>^{i} - X_{0,j}^{i,k}$ and ${\cal
RY}_{j}^{i,k}=<Y_{0,j}>^{i}- Y_{0,j}^{i,k}$.  Ideally, ${\cal
RX}_{j}^{i,k} = {\cal RY}_{j}^{i,k} = 0$ for all $j$ and $k$.  Random
noise causes ${\cal RX}_{j}^{i,k}$ and ${\cal RY}_{j}^{i,k}$ to differ
from zero, but it does not cause any trends with respect to other
quantities.  However, systematic errors can cause such trends.
Anderson \& King (2000) demonstrates that a mismatch between the true
PSF and the ePSF causes ${\cal RX}_{j}^{i,k}$ and ${\cal
RY}_{j}^{i,k}$ to depend on the location of a centroid within a pixel
--- the pixel phase $\Phi_{x}$ or $\Phi_{y}$.  By definition,
$\Phi_{x,j}^{i,k}
\equiv X_{0,j}^{i,k}-Int(X_{0,j}^{i,k})$ and $\Phi_{y,j}^{i,k} \equiv
Y_{0,j}^{i,k}-Int(Y_{0,j}^{i,k})$, where the function $Int(x)$ returns
the integer part of the variable $x$.

	Figure~\ref{fig:rxry-scl1} plots ${\cal RX}$ and ${\cal RY}$
\textit{versus} $\Phi_{x}$ or $\Phi_{y}$ for the SCL~$J0100-3341$
field.  The plots in the panel \ref{rx-ry-9-scl1} are for the 2000
epoch and those in the panel \ref{rx-ry-10-scl1} are for the 2002
epoch.  The filled squares correspond to the QSO and the dots to stars
with a $S/N$ greater than 30.

	The plots of ${\cal RX}$ \textit{versus} $\Phi_{x}$ and ${\cal
RY}$ \textit{versus} $\Phi_{y}$ in Figures~\ref{rx-ry-9-scl1} and
\ref{rx-ry-10-scl1} show trends between these quantities for the QSO.
Values of ${\cal RX}$ and ${\cal RY}$ tend to be negative for
$\Phi_{x}$ and $\Phi_{y}$ less than about 0.5~pixel, and they tend to
be positive for $\Phi_{x}$ and $\Phi_{y}$ greater than about
0.5~pixel.  The points corresponding to the stars do not show these
trends.  The plots of the cross terms, ${\cal RX}$ \textit{versus}
$\Phi_{y}$ and ${\cal RY}$ \textit{versus} $\Phi_{x}$, do not show any
trends for the QSO or for the stars.  These trends indicate a mismatch
between the ePSF and the true PSF (Anderson \& King 2000).  Both stars
with $S/N > 15$ and the QSO contribute to the construction of the
ePSF.  Therefore, the more-extended true PSF of the QSO causes the
ePSF to be wider than an ePSF constructed using only stars; in other
words, the ePSF is a ``compromise'' between that of the stars and that
of the QSO.  An ePSF constructed using objects with $S/N > 100$
diminishes the trends in the values of ${\cal RX}$ and ${\cal RY}$ for
the QSO because the shape of the ePSF is more akin to the shape of the
true PSF of the QSO.  However, increasing the $S/N$ threshold to 100
or more in the construction of the ePSF is undesirable because the
resulting ePSF is poorly sampled because there are only a few stars
with $S/N$ greater than this limit.  Instead, we choose to allow the
errors in the position of the QSO to remain and be reflected in a
greater uncertainty for the measured proper motion for this field.

	Figure \ref{fig:rxry-scl2} plots ${\cal RX}$ and ${\cal RY}$
\textit{versus} $\Phi_{x}$ or $\Phi_{y}$ for the SCL~$J0100-3338$
field.  Figures~\ref{rx-ry-8-scl2}, \ref{rx-ry-9-scl2}, and
\ref{rx-ry-10-scl2} are for the 1999, 2000, and 2002 epochs,
respectively.  Only objects with a $S/N$ greater than 15 are shown.
No plot shows clear evidence for trends between ${\cal RX}$ or ${\cal
RY}$ and $\Phi_{x}$ or $\Phi_{y}$ for the QSO or for the stars.  In this
field, the true PSF of the QSO resembles that for a star, which is
confirmed by Figure~\ref{fig:rf-scl2}, where the values of $\cal{RF}$
for the QSO are indistinguishable from those for stars.

\section{Proper Motion of Sculptor}
\label{sec:pm}

	At this point, there are two lists of fiducial coordinates, one
for each epoch, for the SCL~$J0100-3341$ field, and three for the
SCL~$J0100-3338$ field.  Define the standard coordinate system to be
that which moves uniformly together with the stars of Sculptor.  Thus,
transforming the fiducial coordinates of a star of Sculptor from
different epochs into the standard coordinate system produces the same
value within the measurement uncertainties.  In contrast, the
transformed coordinates of the QSO or any other object that is not a
member of Sculptor will show uniform motion.  The proper motion of
Sculptor derives from the motion of the QSO in the standard coordinate
system.

	P05 describes a procedure for deriving the motion of the QSO,
and any other object that is not a member of the dSph, in the standard
coordinate system from lists of fiducial coordinates at three epochs.
The procedure includes a linear motion in the fitted transformations
between the fiducial coordinate systems and the standard coordinate
system for those objects whose $\chi^2$ calculated with zero motion is
above a threshold.  The SCL~$J0100-3341$ field has only two epochs, so
we have modified the procedure for this case by excluding those
objects with $\chi^2$ values above a threshold from the calculation of
the transformations between the coordinate systems.  The motion of the
QSO is just the difference of the two transformed coordinates.  The
following two sections describe the results from applying these
procedures to the two fields.

\subsection{Motion of the QSO in the SCL~$J0100-3341$ field}
\label{sec:pm1}

	The number of objects with a measured centroid is 567 and 516
in epochs 2000 and 2002, respectively.  Among these, 470 are common to
the two epochs.  The choice for the individual $\chi^2$ that triggers
fitting for uniform linear motion is 15.  The multiplicative constant
that ensures a $\chi^2$ of one per degree of freedom is 1.151 (see
P05 for a discussion of these parameters).

The transformation of the measured centroids to the standard coordinate
system used in this article is
\begin{eqnarray}
x_{j}^{\prime\, i} &=& x_{off} + c_{1} + c_{2}(x_{j}^{i} - x_{off}) +
c_{3}(y_{j}^{i} - y_{off}) 
\label{eq:tranx} \\ 
y_{j}^{\prime\, i} &=& y_{off} + c_{4} + c_{5}(x_{j}^{i} -
x_{off}) + c_{6}(y_{j}^{i} - y_{off})
\label{eq:trany} \\
\sigma_{xj}^{\prime\, i}&=&
\sqrt{(c_{2}\sigma_{xj}^{i})^{2}
+ (c_{3}\sigma_{yj}^{i})^{2}} \\
\sigma_{yj}^{\prime\, i}&=&
\sqrt{(c_{5}\sigma_{xj}^{i})
+ (c_{6}\sigma_{yj}^{i})^{2}}.
\label{eq:tranuy}
\end{eqnarray}
The above represents a modification of the method described in P05,
afforded here because of the greater number of stars.  In the
equations, $c_{1}$ through $c_{6}$ are the free parameters, $(x_{off},
y_{off}) =(512, 512)$~pixel defines the reference point for the
transformation, and $(x^{i}_{j},y^{i}_{j})$ is a measured centroid of
the $i$th object at the $j$th epoch which is transformed to $(x^{\prime
i}_{j},y^{\prime i}_{j})$ in the standard coordinate system.

	Equations 10 and 11 in P05 define position residuals
$RX_{j-1}^{i}$ and $RY_{j-1}^{i}$ for an object $i$ transformed to the
standard coordinate system from the fiducial coordinate system of the
$j$th epoch.  For an ideal case, $RX_{j-1}^{i} = RY_{j-1}^{i} =0$.
Figure~\ref{fig:rx-ry-scl1} shows $RX$ \textit{versus} $X$ and $RY$
\textit{versus} $Y$ for the SCL~$J0100-3341$ field.
The most prominent feature is a ``step'' in $RX_{1-1}$ \textit{versus}
$X$ at $X \simeq 320$~pixel. The values of $RX_{1-1}$ tend to be
negative for $X$ below the step, indicating the presence of a
systematic error in the $X$ coordinates whose source we are unable to
trace.  The values of $RX_{2-1}$ tend to be positive for $X \lesssim
320$~pixel, which is forced by the fitting procedure.

	An \textit{ad hoc} approach for removing the ``steps'' is to
replace $x_{j}^{i}$ with $x_{j}^{i} + c_{7}$ in the
Equations~\ref{eq:tranx} through \ref{eq:tranuy} when $x_{j}^{i} \leq
320$~pixels and to fit for the additional free parameter $c_{7}$.
Applying this remedy removes the ``steps,'' as is shown by
Figure~\ref{fig:rx-ry-scl1-c} which plots the same quantities as
Figure~\ref{fig:rx-ry-scl1}.  In this corrected fitting procedure, the
value of the multiplicative constant that ensures $\chi^2$ of one per
degree of freedom decreased to 1.123 because of the smaller residuals.
The fitted value of $c_7$ is 0.019~pixel.  The proper motion for this
field derives from this fit.  Figure~\ref{fig:rx-ry-scl1-cb} is the
same as Figure~\ref{fig:rx-ry-scl1-c} except that the points are the
weighted mean residuals in ten equal-length bins in $X$ or $Y$.  Note
the different vertical scale.  The points are plotted at the mean of
the coordinate values in the bin.  The average residuals show no
systematic trends above a level of 0.001~pixel.

	Figure~\ref{fig:xq-yq-scl1} shows the location of the QSO as a
function of time in the standard coordinate system.  The top panel
shows the variation of the $X$ coordinate and the bottom panel does
the same for the $Y$ coordinate.  The motion of the QSO is
$(\mu_{x},\mu_{y})=(0.0032 \pm 0.0032, 0.0005 \pm
0.0035)$~pixel~yr$^{-1}$.  The contribution to the total $\chi^2$ from
the QSO, and from any other object whose motion was fit for, is 0
because a line always passes exactly through two points.

\subsection{Motion of the QSO in the SCL~$J0100-3338$ field}
\label{sec:pm2}

	The number of objects with measured centroids is 343, 326, and
314 in epochs 1999, 2000, and 2002, respectively.  Among these, 257
are common to the three epochs.  The choice for the individual
$\chi^2$ that triggers fitting for uniform linear motion is 15.  The
multiplicative constant that ensures a $\chi^2$ of one per degree of
freedom is 1.176.

	Figures~\ref{fig:rx-ry-scl2} and \ref{fig:rx-ry-scl2-b} show
position residuals, $RX$ and
$RY$, as a function of position in the standard coordinate system for
the SCL~$J0100-3338$ field.  They are analogous to
Figures~\ref{fig:rx-ry-scl1} and \ref{fig:rx-ry-scl1-cb}.  From top to bottom, the rows of panels
are for epochs 1999, 2000, and 2002.  No panel shows unambiguous
trends between $RX$ and $X$ or $RY$ and $Y$.  The largest deviations of
the average residuals are $RY \simeq 0.004$~pixel for $Y < 100$~pixel.  Any
systematic trends at the location of the QSO are on the order of
0.001~pixel.  Although not shown in
the figures, the plots of the cross-terms do not show trends either.

	Figure~\ref{fig:xq-yq-scl2} is analogous to
Figure~\ref{fig:xq-yq-scl1} for the SCL~$J0100-3338$ field.  Note that
the slopes in the corresponding plots in Figures~\ref{fig:xq-yq-scl2}\
and \ref{fig:xq-yq-scl1}\ need not be the same because the two fields
are rotated with respect to each other --- though for the fields in
Sculptor the rotation
is only a few degrees.  The uncertainties shown for the points in
Figure~\ref{fig:xq-yq-scl2} are those calculated from the
scatter of the measurements about the mean for an individual epoch
increased by a multiplicative factor.  The introduction of this factor
reduces the contribution to the total $\chi^2$ from the QSO.   Without it,
the contribution was 9.52.  The contribution to
the $\chi^2$ has approximately two degrees of freedom, which implies a
0.9\% probability of a $\chi^2$ larger than 9.52 by
chance.  Such a small probability likely indicates the presence of
unaccounted-for systematic errors.  We choose to increase the uncertainty
in our fitted proper motion by multiplying the uncertainties of the
mean positions at each epoch by the same numerical factor so that contribution
to the total $\chi^2$ is about one per degree of freedom.  Our fitting
procedure calculates a value for the factor for all objects whose
contribution to the total $\chi^2$ exceeds 4.6, which is expected
10\% of the time by chance.  The value of the factor is 2.2 for the
QSO and the uncertainty in the fitted motion of the QSO increases by
essentially the same amount.  The motion of the QSO is
$(\mu_{x},\mu_{y})=(-0.0043\pm 0.0050, 0.0034\pm 0.0038)$~pixel~yr$^{-1}$.

\subsection{Measured Proper Motion}
\label{sec:pmm}

	Table~1 gives the measured proper motion for each field in the
equatorial coordinate system and their weighted mean.  Table~2
tabulates the proper motions for those objects in the SCL~$J0100-3341$
field for which it was measured.  Table~3 does the same for the
SCL~$J0100-3338$ field.  The first line of Table~2 and Table~3
corresponds to the QSO and subsequent objects are listed in order of
decreasing $S/N$.  The ID number of an object is in column~1, the $X$
and $Y$ coordinates of an object in the earliest image of the first
epoch (o65q09010 for SCL~$J0100-3341$ and o5bl02010 for
SCL~$J0100-3338$) are in columns 2 and 3, and the $S/N$ of the object
at the first epoch is in column 4.  The components of the measured
proper motion, expressed in the equatorial coordinate system, are in
columns 5 and 6.  Each value is the measured proper motion in the
standard coordinate system corrected by adding the weighted mean
proper motion of Sculptor given in the bottom line of Table~1.  To
indicate that this correction has been made, the proper motion of the
QSO is given as zero.  The listed uncertainty of each proper motion is
the uncertainty of the measured proper motion, calculated in the same
way as for the QSO, added in quadrature to that of the average proper
motion of the dSph.  The contribution of the object to the total
$\chi^2$ is in column~7.  Although column 7 is in Table 2 for the sake
of symmetry with Table 3, the $\chi^2$ contributions are not
meaningful.

	Schweitzer \etal\ (1995) reports the first measurement of the
proper motion for Sculptor; this study reports an additional two
independent measurements.  Figure~\ref{fig:pm} compares the three
independent measurements, each represented by a rectangle.  A dot at
the center of a rectangle is the best estimate of the proper motion.
The sides of a rectangle are offset from the center by the 1-$\sigma$
uncertainties.  Rectangles 1, 2, and 3 represent the measurements by
Schweitzer \etal (1995), this study (field SCL~$J0100-3341$), and this
study (field SCL~$J0100-3338$), respectively.

	The $\alpha$ components of our measurements 2 and 3 agree
almost exactly and their $\delta$ components differ by only
1.4$\times$ the uncertainty of their difference.  While the
$\delta$ component of measurement 1 agrees with the $\delta$
components of measurements 2 and 3, the $\alpha$ component does not
agree with either one.  The $\alpha$ components of measurements 1 and
2 differ by 2.3$\times$ the uncertainty of their difference and those
for measurements 1 and 3 differ by 2.2$\times$.  Because of the large
difference in the $\alpha$ components of the proper motion between the
measurement from Schweitzer \etal\ (1995) and from our two fields, we
choose to use the weighted average proper motion from Table~1 to
determine the space velocity of Sculptor.

\subsection{Galactic Rest Frame Proper Motion}
\label{sec:pmgrf}

	The measured proper motion of the dSph contains contributions
from the motion of the LSR and the peculiar motion of the Sun.  The
magnitude of the contributions depend on the Galactic longitude and
latitude of the dSph.  Removing them yields the Galactic-rest-frame
proper motion --- the proper motion measured by a hypothetical
observer at the location of the Sun but at rest with respect to the
Galactic center.  Columns (2) and (3) of Table~4 give the equatorial
components, $(\mu_{\alpha}^{\mbox{\tiny{Grf}}},
\mu_{\delta}^{\mbox{\tiny{Grf}}})$, of the Galactic-rest-frame proper
motion.  Their derivation assumes: 220~km~s$^{-1}$ for the circular
velocity of the LSR; 8.5~kpc for the distance of the Sun from the
Galactic center; and $(u_\odot, v_\odot, w_\odot) = (-10.00 \pm 0.36,
5.25 \pm 0.62 , 7.17 \pm 0.38)$~km~s$^{-1}$ (Dehnen \& Binney 1998)
for the peculiar velocity of the Sun, where the components are
positive if $u_{\odot}$ points radially away from the Galactic center,
$v_{\odot}$ is in the direction of rotation of the Galactic disk, and
$w_\odot$ points in the direction of the North Galactic Pole.  Columns
(4) and (5) give the Galactic-rest-frame proper motion in the Galactic
coordinate system,
$(\mu_{l}^{\mbox{\tiny{Grf}}},\mu_{b}^{\mbox{\tiny{Grf}}})$.  The next
three columns give the $\Pi$, $\Theta$, and $Z$ components of the
space velocity in a cylindrical coordinate system centered on the
dSph.  The components are positive if $\Pi$ points radially away from
the Galactic axis of rotation, $\Theta$ points in the direction of
rotation of the Galactic disk, and $Z$ points in the direction of the
North Galactic Pole.  The derivation of these components assumes
87~kpc (Kaluzny
\etal\ 1995) for the heliocentric distance to and $109.9 \pm
1.4$~km~s$^{-1}$ (Queloz, Dubath, \& Pasquini 1995) for the
heliocentric radial velocity of Sculptor.  The last two columns give
the radial and tangential components of space velocity for an observer
at rest at the Galactic center.  The component $V_{r}$ is positive if
it points radially away from the Galactic center.  Thus, at present,
Sculptor is moving away from the Milky Way.

\section{Orbit and Orbital Elements of Sculptor}
\label{sec:orbit}

	Knowing the space velocity of a dSph permits a determination
of its orbit for a given form of the Galactic potential.  This study
adopts a Galactic potential that has a contribution from a disk of the
form (Miyamoto \& Nagai 1975)
\begin{equation}
\label{diskpot}
\Psi_{\mbox{\small{disk}}}=-\frac{G
M_{\mbox{\small{disk}}}}{\sqrt{R^{2}+(a+\sqrt{Z^{2}+b^{2}})^{2}}},
\end{equation}
from a spheroid of the form (Hernquist 1990)
\begin{equation}
\label{spherpot}
\Psi_{\mbox{\small{spher}}}=-\frac{GM_{\mbox{\small{spher}}}}
{R_{\mbox{\small{GC}}}+c},
\end{equation}
and from a halo of the form
\begin{equation}
\label{logpot}
\Psi_{\mbox{\small{halo}}}=v^{2}_{\mbox{\small{halo}}}\ln
(R^{2}_{\mbox{\small{GC}}}+d^{2}).
\end{equation}
In the above equations, $R_{\mbox{\small GC}}$ is the Galactocentric
distance, $R$ is the projection of $R_{\mbox{\small GC}}$ onto the
plane of the Galactic disk, and $Z$ is the distance from the plane of
the disk.  All other quantities in the equations are adjustable
parameters and their values are the same as those adopted by Johnston,
Sigurdsson, \& Hernquist (1999):
$M_{\mbox{disk}}=1.0\times10^{11}$~M$_{\odot}$,
$M_{\mbox{spher}}=3.4\times10^{10}$~M$_{\odot}$,
$v_{\mbox{halo}}=128$~km~s$^{-1}$, $a=6.5$~kpc, $b=0.26$~kpc,
$c=0.7$~kpc, and $d=12.0$~kpc.

	Figure~\ref{fig:orbit} shows the projections of the orbit of
Sculptor onto the $X-Y$ (top-left panel), $X-Z$ (bottom-left panel),
and $Y-Z$ (bottom-right panel) Cartesian planes.  The orbit results
from an integration of the motion in the Galactic potential given by
Equations~\ref{diskpot}, \ref{spherpot}, and \ref{logpot}.  The
integration extends for 3~Gyr backwards in time and begins at the
current location of Sculptor with the negative of the space velocity
components given in the bottom line of columns (6), (7), and (8) of
Table~4.  The filled square marks the current location of the dSph, the
filled star indicates the center of the Galaxy, and the two small
circles mark the points on the orbit where $Z=0$ or, in other words,
where the orbit crosses the plane of the Galactic disk.  The large
circle is for reference: it has a radius of 30~kpc.  In the
right-handed coordinate system of Figure~\ref{fig:orbit}, the current
location of the Sun is on the positive $X$-axis.  The figure shows that
Sculptor is moving away from the Milky Way, is closer to perigalacticon
than apogalacticon, and that it has a nearly polar orbit with a modest
eccentricity.

	Table~5 tabulates the elements of the orbit of Sculptor.  The
value of the quantity is in column (4) and its $95\%$ confidence
interval is in column (5).  The latter comes from 1000 Monte Carlo
experiments, where an experiment integrates the orbit using an initial
velocity that is generated by randomly choosing the line-of-sight
velocity and the two components of the measured proper motion from
Gaussian distributions whose mean and standard deviation are the best
estimate of the quantity and its quoted uncertainty, respectively.  The
eccentricity of the orbit is defined as
\begin{equation}
\label{eccentricity}
e = \frac{(R_{a} - R_{p})}{(R_{a} + R_{p})}.
\end{equation}
The most likely orbit has about a 2:1 ratio of apogalacticon to
perigalacticon, though the 95\% confidence interval for the
eccentricity allows ratios approximately between 1.7:1 and 4:1.  The
orbital period of Sculptor, 2.2~Gyr, is about 50\% longer than those of
Carina (1.4~Gyr; P03) and Ursa~Minor (1.5~Gyr; P05).

\section{Discussion}
\label{sec:disc}

	Knowing the orbit can help answer several questions about
Sculptor, or, at least, increase the level of our understanding of
this galaxy.  These questions are: 1. Is Sculptor a member of a stream
of galaxies? 2. Is its star formation history correlated with the
orbit?  3. What is the origin of the HI clouds detected in close
proximity to the dSph?  4. Does Sculptor contain dark matter?

\subsection{Is Sculptor a Member of a Stream?}
\label{stream}

	Lynden-Bell \& Lynden-Bell (1995) proposes that Sculptor may be
a member of one of three possible streams: stream No.~2 (together with
the LMC, SMC, Draco, Ursa Minor, and Carina); No.~4a (together with
Sextans, Pal 3, and Fornax); or No.~4b (together with Sextans and
Fornax).  Columns (2) and (3) in Table~6 give the predicted
heliocentric (i.e., ``measured'' in our terminology) proper motion in
the equatorial coordinate system for Sculptor if it indeed belongs to
any of the three streams.  The magnitude of the proper motion vector,
$\vert \vec{\mu} \vert = \sqrt {\mu_{\alpha}^{2} + \mu_{\delta}^{2}}$,
and its position angle are in columns (4) and (5).  For easy
comparison, the corresponding quantities from our study are in the
bottom line of the table.  Comparing the entries shows that the
predictions for streams 2 and 4a disagree significantly with our
measurement.  However, the prediction for stream No. 4b is closer: the
magnitudes differ by 1.6$\times$ the uncertainty in their difference,
while the position angles differ by 1.6$\times$.  Differences of this
size should occur by chance 1\% of the time.  The measured proper
motion based on only the three-epoch data in the SCL~$J0100-3338$ field
improves the agreement with the prediction for stream 4b.  Thus, while
we rule out the possibility that Sculptor is a member of stream No.~2
or 4a, its membership in stream~4b is possible.

	Stream 4b contains both Sculptor and Fornax.  The Dinescu
\etal\ (2004) proper motion for Fornax is $(\mu_\alpha, \mu_\delta) =
(59 \pm 16, -15 \pm 16)$~mas~cent$^{-1}$.  The magnitude and position
angle of the proper motion are $61 \pm 16$~mas~cent$^{-1}$ and $104\pm
15$~degrees.  The prediction for stream 4b from Lynden-Bell \&
Lynden-Bell (1995) is 20~mas~cent$^{-1}$ and 162~degrees.  The
difference between the measured and predicted proper motions would be
this large or larger by chance only 0.4\% of the time.  Thus, the
physical reality of stream~4b is doubtful.

	Dinescu \etal\ (2004) argues that Fornax and Sculptor are
members of the same stream that also includes Leo I, Leo II, and
Sextans.  Together the galaxies define the FL$^{2}$S$^{2}$ plane.  If
they do form a stream, their Galactic-rest-frame proper motion vectors
should be aligned with the great circle passing through the galaxies.
The position angle of the great circle passing through Sculptor and
Fornax is about 99~degrees at the location of Sculptor and 95~degrees
at Fornax.  The position angle of the Galactic-rest-frame proper
motion for Fornax reported by Dinescu \etal\ (2004) is $79 \pm
25$~degrees, which differs by 0.64$\times$ its uncertainty from the
position angle of the great circle.  If Sculptor and Fornax form a
stream, then they should move in the same direction along the great
circle connecting them.  Thus, the proper motion of Fornax from Dinescu
\etal\ (2000) implies that the position angle of the
Galactic-rest-frame proper motion of Sculptor should be 99~degrees.
The position angle for the proper motion of Sculptor reported here is
$333 \pm 15$~degrees, which differs from the
prediction by 8.4$\times$ its uncertainty.  Discounting the
proper motion from this study and instead using the position angle of
$40 \pm 24$~degrees implied by the proper motion measured by
Schweitzer \etal\ (1995) does not remove the disagreement: the
position angle differs by $2.5\times$ its uncertainty from that of the
great circle.  We conclude that Sculptor and Fornax do not belong to
the same stream.

	Kroupa \etal\ (2004) shows that the 11 dwarf
galaxies nearest to the Milky Way are nearly on a plane, whose two
poles are at $(\ell,b) = (168,-16)$~degrees and $(348,+16)$~degrees.
Adopting the direction of the angular momentum vector as the pole of
the orbit, then the location of the pole is
\begin{equation} 
(\ell,b) = (\Omega+90^{\circ},\Phi-90^{\circ}).  
\end{equation} 
Because of the left-handed nature of the Galactic rotation, prograde
orbits have $b < 0$ and retrograde orbits have $b > 0$.  Thus, the pole
of our orbit for Sculptor is $(\ell,b) = (5 \pm 16, -4 \pm
1.6)$~degrees, where the uncertainties are 1-$\sigma$ values from the
Monte Carlo simulations.  The galactic longitudes of the poles of the
plane and orbit agree within the uncertainty, but the galactic
latitudes do not.  They differ by 20~degrees, which is more than
12$\times$ the uncertainty in the location of the pole of the orbit.
However, there is also some uncertainty in the orientation of the plane
passing through the dwarf galaxies near the Milky Way.  We conclude the
plane of the orbit of Sculptor is similar to the plane defined by the
nearby dwarf galaxies.

\subsection{The Effect of the Galactic Tidal Force on the Structure of
Sculptor}
\label{sec:tides}

	The measured ellipticity of the isodensity contours increases
with projected distance from the center of Sculptor (see Figure~1 in
Irwin \& Hatzidimitriou 1995), akin to surface density contours of a
model dSph in the numerical simulations of Johnston, Spergel, \&
Hernquist (1995; e.g., see Figure~4).  If the Galactic tidal force
deformed Sculptor from an initial nearly-spherical shape to its
present elongated shape in the outer regions then, from our vantage
point nearly in the orbit plane, the position angle of its projected
major axis should be similar to --- or differ by 180 degrees from ---
the position angle of the Galactic-rest-frame proper motion vector, as
predicted by the numerical simulations of Oh, Lin, \& Aarseth (1995),
Piatek \& Pryor (1995), or Johnston, Spergel, \& Hernquist (1995).
The position angle of the projected major axis is $99 \pm 1$~degrees
and the position angle of our measured Galactic-rest-frame proper
motion vector is $333 \pm 15$~degrees.  Allowing for the
180-degree degeneracy, the difference between the two position angles
is 3.6$\times$ the uncertainty of their difference, which
suggests that the Galactic tidal force has not elongated Sculptor.

\subsection{Does Star Formation History Correlate with the Orbital
Motion of Sculptor?}
\label{sec:sfh}

	Da Costa (1984) imaged a 3~arcmin $\times$ 5~arcmin field
located just outside of the core radius of Sculptor in three bands:
$B$, $V$, and $R$.  The photometry reaches the main-sequence turn off.
Comparing theoretical isochrones with the distribution of stars in the
color-magnitude diagram, the study finds that the majority of stars is
about 2-3 Gyr younger than the galactic globular clusters of comparable
metal abundance.  An earlier study by Kunkel \& Demers (1977) based on
$B$ and $V$ photometry extending to 0.4 magnitudes below the horizontal
branch reaches a similar conclusion.  The color-magnitude diagram also
shows a population of ``blue stragglers,'' which might be indicative of
an extended period of star formation.  Da Costa (1984) concludes,
however, that, if an intermediate-age stellar population exists in
Sculptor, it is ``infinitesimal compared to that of the Carina system.''

	Deep \textit{HST} imaging by Monkiewicz \etal\ (1999) in a
single field, reaching 3 magnitudes below the main-sequence turn-off,
confirms the basic picture of Sculptor uncovered by Kunkel \& Demers
and Da Costa (1984).  This color-magnitude diagram also reveals the
presence of ``blue stragglers'' and implies an age comparable to that
of the galactic globular clusters.  The small number of stars in the
small field made a search for an intermediate-age stellar population
inconclusive.

	Majewski \etal\ (1999), Hurley-Keller \etal\ (1999), and
Harbeck \etal\ (2001) use wide-field imaging to show the presence of
two stellar populations with distinctly different metallicities
([Fe/H] = --2.3 and --1.5; Majewski \etal\ 1999).  The more metal rich
population is more centrally concentrated in the galaxy.  Most
recently, Tolstoy \etal\ (2004) confirms the above picture using
wide-field imaging and spectroscopy.  Spectroscopically-determined
metallicities range from --2.8 to --0.9.  Stars more metal-rich than
--1.7 are more centrally concentrated and have a smaller velocity
dispersion than the rest of the sample.  However, both stellar
populations are older than 10~Gyrs.

	The aforementioned studies show that there were at least two
episodes of star formation at times more than 10~Gyr ago.  Because
10~Gyrs is much longer than the orbital period of approximately
2.2~Gyr, there is no clear connection between the stimulation of
star formation and processes such as the Galaxy-Sculptor tidal
interaction or the effects of ram pressure.  The lack of correlation
could be due to the loss of all of the gas in Sculptor about 10~Gyr
ago.  But, surprisingly, the observations indicate that Sculptor has
HI today.

\subsection{HI Gas in Sculptor}
\label{sec:gas}

	Unlike most other Galactic dSphs, Sculptor contains a
detectable amount of HI.  Knapp \etal\ (1978) detects three clouds of
HI in the vicinity of Sculptor and speculates that one of them, with a
radial velocity of 120~km~s$^{-1}$, may be associated with Sculptor,
whose radial velocity at the time was uncertain.  Carignan
\etal\ (1998) confirms and refines this detection and puts a lower
limit on the mass of HI of $3.0 \times 10^{4}$~M$_{\odot}$.  Bouchard
\etal\ (2003) repeats the observations over a wider field with the
Parkes single-dish telescope and, over a smaller region, at higher
angular resolution with the Australia Telescope Compact Array.  The
better data show that the HI is not associated with a background
galaxy and the probability of a chance superposition of a galactic
high velocity cloud is less than 2\%.  These arguments, together with
the agreement within 4~km~s$^{-1}$ of the radial velocity of the HI
and the radial velocity of Sculptor make a strong case for the
physical association of the clouds with the dSph.

	The gas is in two clouds.  Figure~\ref{fig:gas}\ shows the
distribution of HI on the sky in the direction of Sculptor based on
the Australia Telescope Compact Array data from Bouchard \etal\
(2003).  The asterisk marks the optical center of Sculptor and the two
clouds are about 20--30~arcmin from the center, one to the northeast
and one to the southwest.  The masses of the clouds are $(4.1 \pm 0.2)
\times 10^4$~M$_\odot$ and $(1.93 \pm 0.02) \times 10^5$~M$_\odot$,
respectively.  The two clouds lie nearly along the minor axis of the
dSph; the orientation of Sculptor is shown in Figure~\ref{fig:gas}\ by
the ellipse representing the optical boundary.

	Although the association between the gas and the dSph seems
well-established, why Sculptor still has gas when most of the other
galactic dSphs do not and the cause of the observed configuration of
the gas are debated.  Mayer \etal\ (2005) studies the loss of gas from
dwarf galaxies in the Local Group using numerical simulations that
include ram pressure stripping.  It quantifies the expected result that
a galaxy with a deeper potential well or with a larger perigalacticon
is more likely to retain its gas.  The orbit of Sculptor is similar to
those of Carina and Ursa Minor, which suggests that differences in the
degree of tidal shocking or ram pressure stripping are not the reason
for the difference in gas retention.  However, note that the 95\%
confidence intervals for the perigalacticons of all three orbits are
still too large to make a conclusive statement.  The larger mass of
Sculptor compared to Carina and Ursa Minor is the most likely reason
why it was able to retain gas.

	Mechanisms that could affect the distribution of HI within the
dSph are tidal interaction, ram pressure, and forces from supernovae
or winds from young stars.  Tidal interaction and ram pressure tend to
spread the gas in the plane of the orbit.  The two arrows in
Figure~\ref{fig:gas}\ represent the Galactic-rest-frame proper motions
as measured by Schweitzer \etal\ (1995; dashed) and this study
(solid).  The ratio of their lengths is the same as the ratio of the
magnitudes of the proper motions.  The dotted line is a section of a
great circle that passes through Sculptor and Fornax; its position
angle is 99~degrees.

	The Schweitzer \etal\ (1995) proper motion (dashed arrow) is
nearly aligned with the line that connects the centers of the clouds.
Carignan \etal\ (1998) notes this alignment and suggests that it may
indicate a ``tidal'' origin for the clouds: presumably, the Galactic
tidal force stretches the gas, initially centered within the dSph, into
its observed distribution.  The most serious problem with such a
picture is that the tidal force would stretch the distribution of both
the stars and the gas, which then aligns the major axes of the gaseous
and stellar distributions from our perspective as an observer nearly in
the plane of the orbit of Sculptor.  They are not aligned, perhaps
because the motion of the gas is governed by both gravitational forces
and pressure gradients.  As was noted in Section~\ref{sec:tides}, if
the observed ellipticity of Sculptor is due to the tidal force, then
the Galactic-rest-frame proper motion should be aligned with the major
axis.  Figure~\ref{fig:gas}\ shows that the solid arrow, our
measurement, is closer to such an alignment than the dashed arrow.
Because the tidal force is zero at the center of a dSph, it cannot by
itself separate a single cloud centered on the dSph into two clouds.
Thus, within the context of a picture in which tides have had an
important effect on Sculptor, our proper motion is more plausible than
that of Schweitzer \etal\ (1995).  However, is our proper motion
consistent with the geometry of the two clouds?

	We think yes.  First, the two clouds are elongated in the
direction of our proper motion (see particularly Figure~1 in Bouchard
\etal\ 2003).  The observed elongation could be due to ram pressure
from the motion through a gaseous Galactic halo.  Second and more
speculatively, the two clouds could be due to the Rayleigh-Taylor
instability when the HI in Sculptor moves through the hot and
low-density gaseous halo.  Or the gas could have been squeezed out
perpendicular to the direction of motion by the compressive tidal
shock when Sculptor crosses the Galactic disk.  Expansion combined
with infall of the gas forms a ring that looks like two clouds in
projection.

\subsection{Is there dark matter in Sculptor?}
\label{sec:dm}

	Estimates of the $M/L_V$ for Sculptor range from $6 - 13$ and
estimates of the limiting radius range from about $40 - 80$~arcmin.
The estimates of $M/L$ assume that mass follows light.  If this is
true, then the implied mass of Sculptor must be large enough, given our
orbit, to produce a tidal radius that is at least as large as the
observed limiting radius.  Equating the tidal radius and the limiting
radius predicts a value for $M/L$, which should agree with the measured
value.  Also, the dSph must have a mass and, hence, $M/L$ large enough
for it to have survived destruction by the Galactic tidal force on our
orbit.

In lieu of numerical simulations, an approximate analytical approach is
to calculate the tidal radius, $r_t$, beyond which a star becomes
unbound from the dSph.  For a logarithmic Galactic potential, $r_t$ is
given by (King 1962; Oh, Lin, \& Aarseth 1992)
\begin {equation}
r_t = \left(\frac{(1-e)^2}{[(1+e)^2/2e]\ln[(1+e)/(1-e)] +1} \,
\frac{M}{M_G}\right)^{1/3} a.
\label{eq:rtidal}
\end {equation}
Here $e$ is eccentricity of the orbit, $a$ is the semi-major axis ($a
\equiv (R_{a}+R_{p})/2$), $M$ is the mass of the dSph, and $M_G$ is the
mass of the Galaxy within $a$.  Equating $r_t$ with the observed
limiting radius derived by fitting a King (1966) model, $r_k$, gives a
value for $M/L_V$ for a given orbit.  If $r_k = 40$~arcmin, then 28\%
of the orbits in Monte Carlo simulations have $M/L_V > 6$ and
10\% of the orbits have $M/L_{V} > 13$. If $r_k = 80$~arcmin, then
100\% of the orbits have $M/L_V > 13$.  These
results show that the global $M/L$ of Sculptor is probably larger than
the measured $M/L$, if the larger of the measured limiting radii is
identified with the
tidal radius.  The $M/L$ calculated assuming that mass follows light
underestimates the true global $M/L$ if Sculptor contains dark matter
that is more spatially extended than the luminous matter (e.g., Pryor
\& Kormendy 1990).  However, equation~\ref{eq:rtidal}\ shows that $M
\propto r_t^3$, so the values for $M/L$ derived using this equation
are sensitive to the measured value of the limiting radius.  Until
kinematic measurements definitively identify the tidal radius, an
$M/L$ derived with the above argument should be treated with caution.

	The average of the measured values of $M/L_V$ for Galactic globular
clusters is 2.3 (Pryor \& Meylan 1993).  Could the true $M/L_V$ of
Sculptor be similar to this average?  Numerical simulations by Oh, Lin,
\& Aarseth (1995) and Piatek \& Pryor (1995) show that the ratio of the
limiting radius derived by fitting a theoretical King model (King
1966), $r_k$, to the tidal radius defined by Equation~(\ref{eq:rtidal})
is a useful indicator of the importance of the Galactic tidal force on
the structure of a dSph.  These simulations show that: if $r_{k}/r_t
\lesssim 1.0$, the Galactic tidal force has little effect on the
structure of the dSph; at $r_k/r_t \approx 2.0$, the effect of the
force increases rapidly with increasing $r_k/r_t$; and, for $r_k/r_t
\approx 3.0$, the dSph disintegrates in a few orbits.  Assuming that
$M/L_V = 2.3$ and $r_k = 40$~arcmin, $r_k/r_t > 2.0$ for 6\%
of the orbits generated in Monte Carlo simulations.  If $r_k = 80$~arcmin,
the fraction is 100\%.  Thus, it is possible
that Sculptor could have survived for a Hubble time on its current
orbit if it only contains luminous matter.  We conclude that measured
orbit of Sculptor does not require it to contain dark matter.

\subsection{A Lower Limit for the Mass of the Milky Way}
\label{sec:massofg}

	Sculptor is bound gravitationally to the Milky Way.  The
Galactocentric space velocity of the dSph imposes a lower limit on the
mass of the Milky Way within the present Galactocentric radius of the
dSph, $R$.  Assuming a spherically symmetric mass distribution and
zero for the total energy of the dSph, the lower limit for the mass of
the Milky Way is given by
\begin{equation}
\label{mwmass}
M=\frac{R\left (V_{r}^{2} + V_{t}^{2} \right )}{2G}.
\end{equation}
Setting $R=87$~kpc and using the values from Table~4 for $V_{r}$ and
$V_{t}$, $M = (4.6\pm 2.0) \times 10^{11}\ M_{\odot}$.
This lower limit is consistent with other recent estimates of the mass
of the Milky Way, such as the mass of $5.4^{+0.1}_{-0.4} \times
10^{11}\ M_{\odot}$ within $R=50$~kpc found by Sakamoto, Chiba, \&
Beers (2003).  The Milky Way potential adopted in
Section~\ref{sec:orbit} has a mass of $7.8\times 10^{11}\ M_{\odot}$
out to $R=87$~kpc.

\section{Summary}
\label{sec:sum}

	This article presents a measurement of the proper motion of
Sculptor using data taken with \textit{HST} and STIS in imaging mode.
Using this measurement, it derives the orbit and discusses: membership
in proposed streams, tidal interaction with the Milky Way, the relation
between the orbit and the star-formation history, the HI gas associated
with the dSph, the dark matter content, and a lower limit on the mass
of the Milky Way.  The list below enumerates our findings.

	1. Two independent measurements of the proper motion give a
weighted mean value of $(\mu_\alpha,\mu_\delta)=(9\pm 13, 2 \pm
13)$~mas~cent$^{-1}$ in the equatorial coordinate system for a
heliocentric observer.

	2. Removing the contributions to the measured proper motion
from the motions of the Sun and of the LSR gives a Galactic-rest-frame
proper motion of $(\mu_{\alpha}^{\mbox{\tiny{Grf}}},
\mu_{\delta}^{\mbox{\tiny{Grf}}})=(-23\pm 13, 45 \pm 13)$~mas~cent$^{-1}$
in the equatorial coordinate system for an
observer at the location of the Sun but at rest with respect to the
Galactic center.  In the Galactic coordinate system this motion is
$(\mu_{l}^{\mbox{\tiny{Grf}}},\mu_{b}^{\mbox{\tiny{Grf}}}) = (11
\pm 13, -50 \pm 14)$~mas~cent$^{-1}$.

	3. The radial and tangential components of the space velocity
are $V_{r}=79 \pm 6$~km~s$^{-1}$ and $V_{t}=198 \pm 50$~km~s$^{-1}$, respectively, as measured by a Galactocentric
observer at rest.

	4. The best estimate of the orbit shows that Sculptor is
approaching its apogalacticon of $R_{a}=122$~kpc on a polar
orbit with eccentricity $e=0.29$.  The perigalacticon of the orbit is
$R_{p}=68$~kpc and the orbital period is $T=2.2$~Gyr.

	5. Sculptor is not a member of streams~2 and 4a proposed by
Lynden-Bell \& Lynden-Bell (1995).  It could be a member of stream~4b,
though the proper motion of Fornax measured by Dinescu \etal\ (2004)
makes the physical reality of this stream doubtful.

        6. The proper motion vectors of Sculptor and Fornax show that
they cannot be members of the same stream.

	7. The pole of the orbit of Sculptor is 26~degrees from the
pole of the plane of the Galactic dSphs noted by Kroupa \etal\ (2004).
This difference is much larger than the uncertainty in the pole of the
orbit, but is probably within the uncertainty of the definition of the
plane of the dSphs.

	8. A comparison of the orbit of Sculptor to those of other
dSphs does not provide a clear reason for why Sculptor contains HI
while the others do not.  The origin and distribution of HI remain
puzzling.  This article proposes that, while the line connecting the
two clouds of HI is nearly perpendicular to our Galactic-rest-frame
proper motion, some combination of ram pressure, tidal interaction, and
Rayleigh-Taylor instability could produce this geometry.

\acknowledgments

We thank the anonymous referee for comments that helped to improve the
presentation of our work.  We also thank Sergei Maschenko for pointing
out to us that our method for propagating uncertainties from the
measured proper motions to the space velocity was incorrect.  CP and SP
acknowledge the financial support of the Space Telescope Science
Institute through the grants HST-GO-07341.03-A and HST-GO-08286.03-A
and from the National Science Foundation through the grant
AST-0098650.  EWO acknowledges support from the Space Telescope Science
Institute through the grants HST-GO-07341.01-A and HST-GO-08286.01-A
and from the National Science Foundation through the grants AST-9619524
and AST-0098518.  MM acknowledges support from the Space Telescope
Science Institute through the grants HST-GO-07341.02-A and
HST-GO-08286.02-A and from the National Science Foundation through the
grant AST-0098661.  DM is supported by FONDAP Center for Astrophysics
15010003.

\setcounter{figure}{0}
\clearpage
\begin{figure}
\includegraphics[scale=.7,angle=-90]{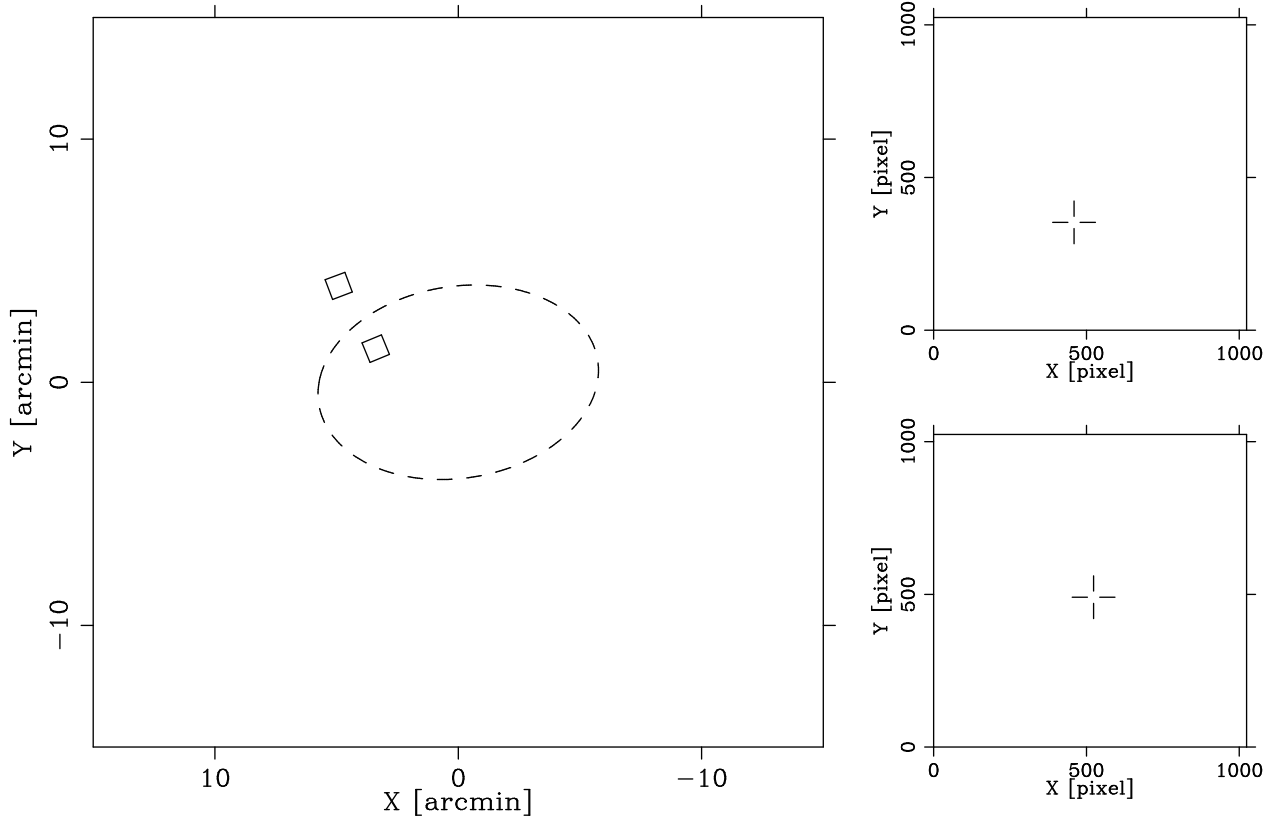}
\caption{Left panel: An image of the sky in the direction of the
Sculptor dSph.  The dashed ellipse has an ellipticity of 0.32 and a
semi-major axis equal to the Irwin \& Hatzidimitriou (1995) core
radius.  The two squares represent the fields studied in this article.
The one within the core corresponds to the SCL~$J0100-3341$ field and
the one outside to the SCL~$J0100-3338$ field.  Top-right panel: A
sample image from the epoch 2000 data for the SCL~$J0100-3341$ field.
The cross-hair indicates the location of the QSO.  Bottom-right panel:
A sample image from the epoch 1999 data for the SCL~$J0100-3338$
field.  Again, the cross-hair indicates the location of the QSO.}
\label{fig:fields}
\end{figure}

\clearpage
\begin{figure}
\includegraphics[scale=.5,angle=-90]{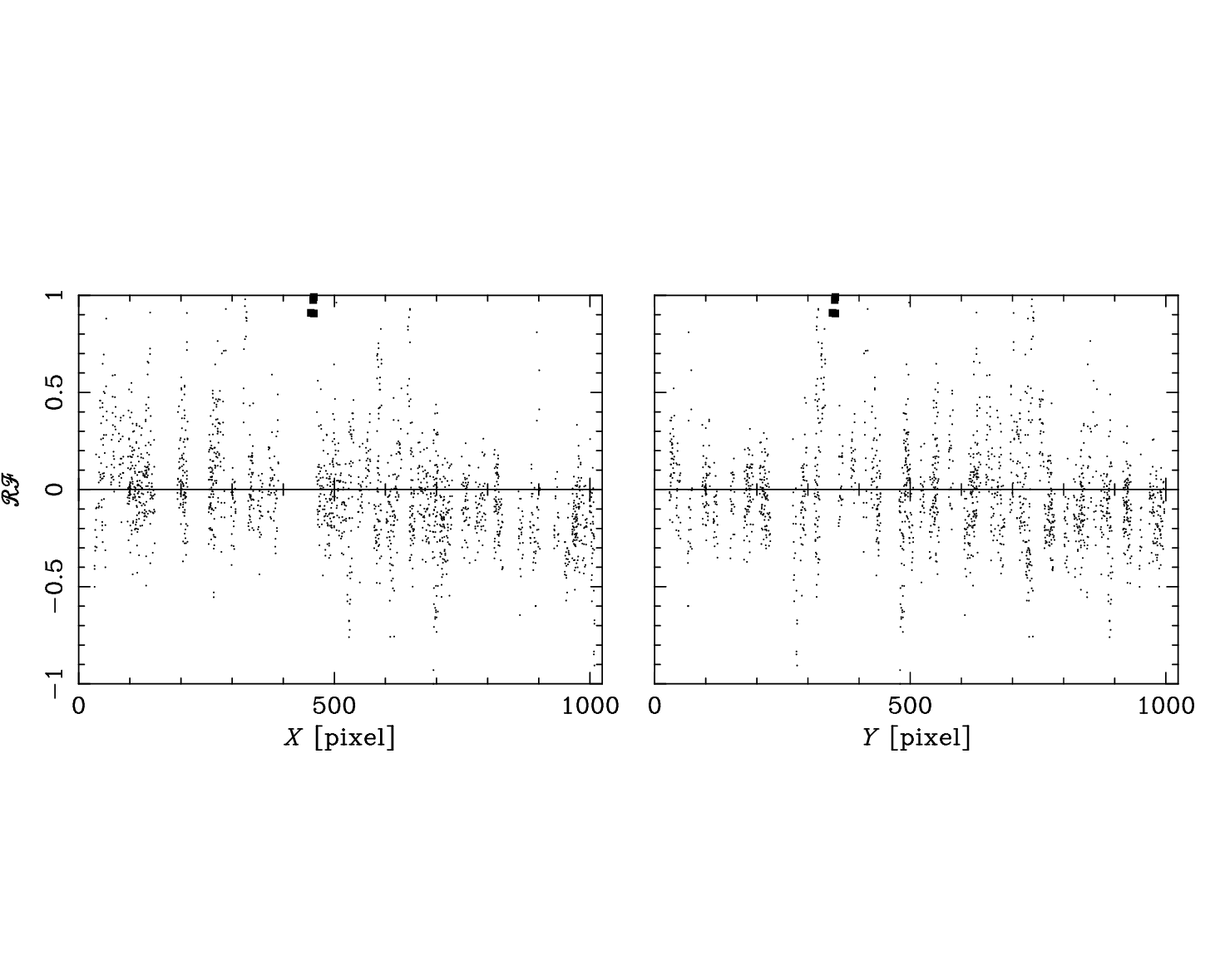}\\
\includegraphics[scale=.5,angle=-90]{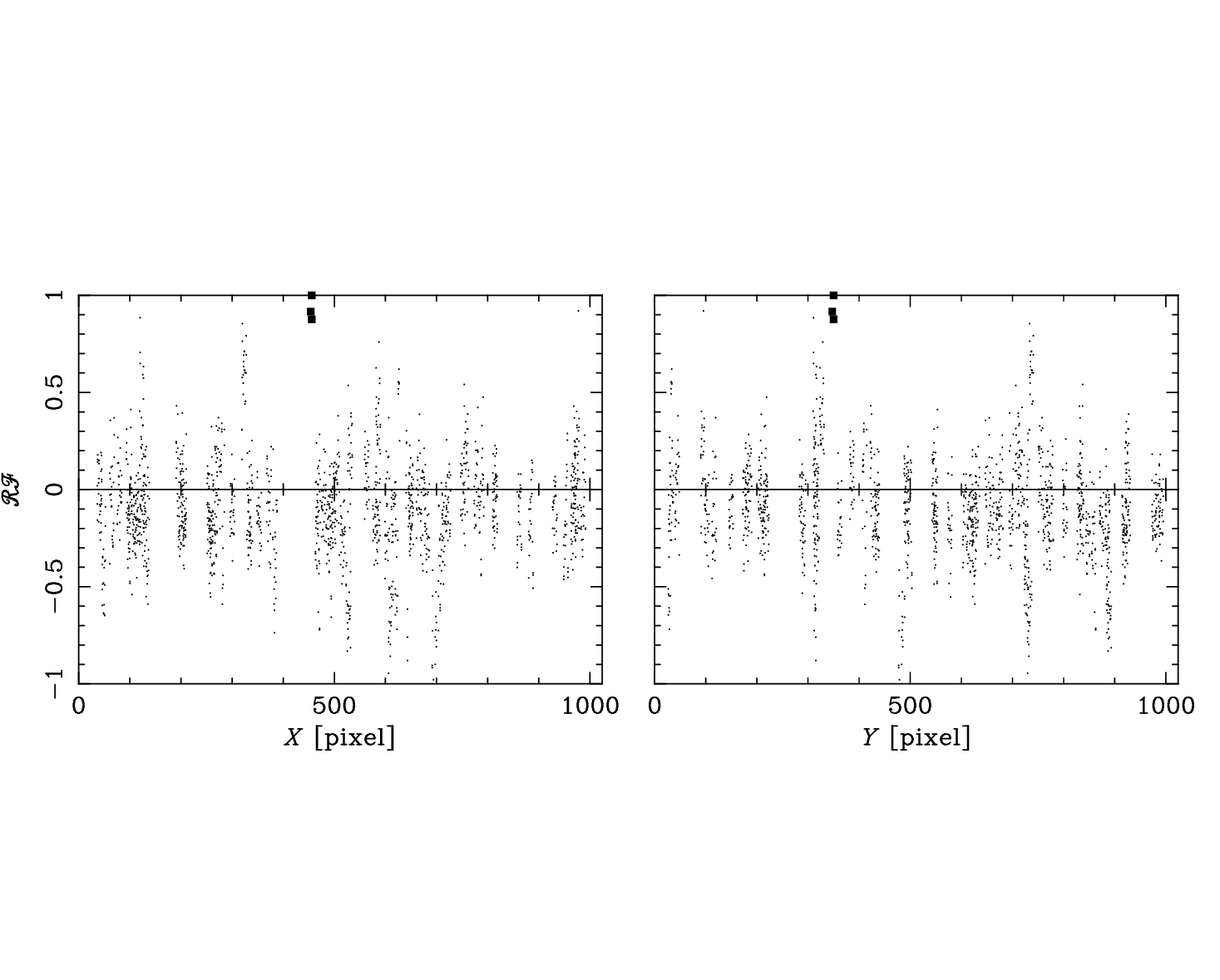}
\caption{Flux residual \textit{versus} location for objects in an image
for the SCL~$J0100-3341$ field.  Every plot displays points from 24
images.  The left-hand panels plot ${\cal RF}$ \textit{versus} the
$X$-coordinate and the right-hand panels plot ${\cal RF}$
\textit{versus} the $Y$-coordinate.  The filled squares represent the
QSO.  The top row of panels corresponds to the 2000 epoch and the
bottom row to the 2002 epoch.  For ease of comparison, all of the
plots have the same scale on the vertical axis.  Only objects with a
$S/N$ greater than 30 are plotted.}
\label{fig:rf-scl1}
\end{figure}

\clearpage
\begin{figure}
\centering
\subfigure[]
{\label{stis-8-rf-x-2}
\includegraphics[scale=.32,angle=-90]{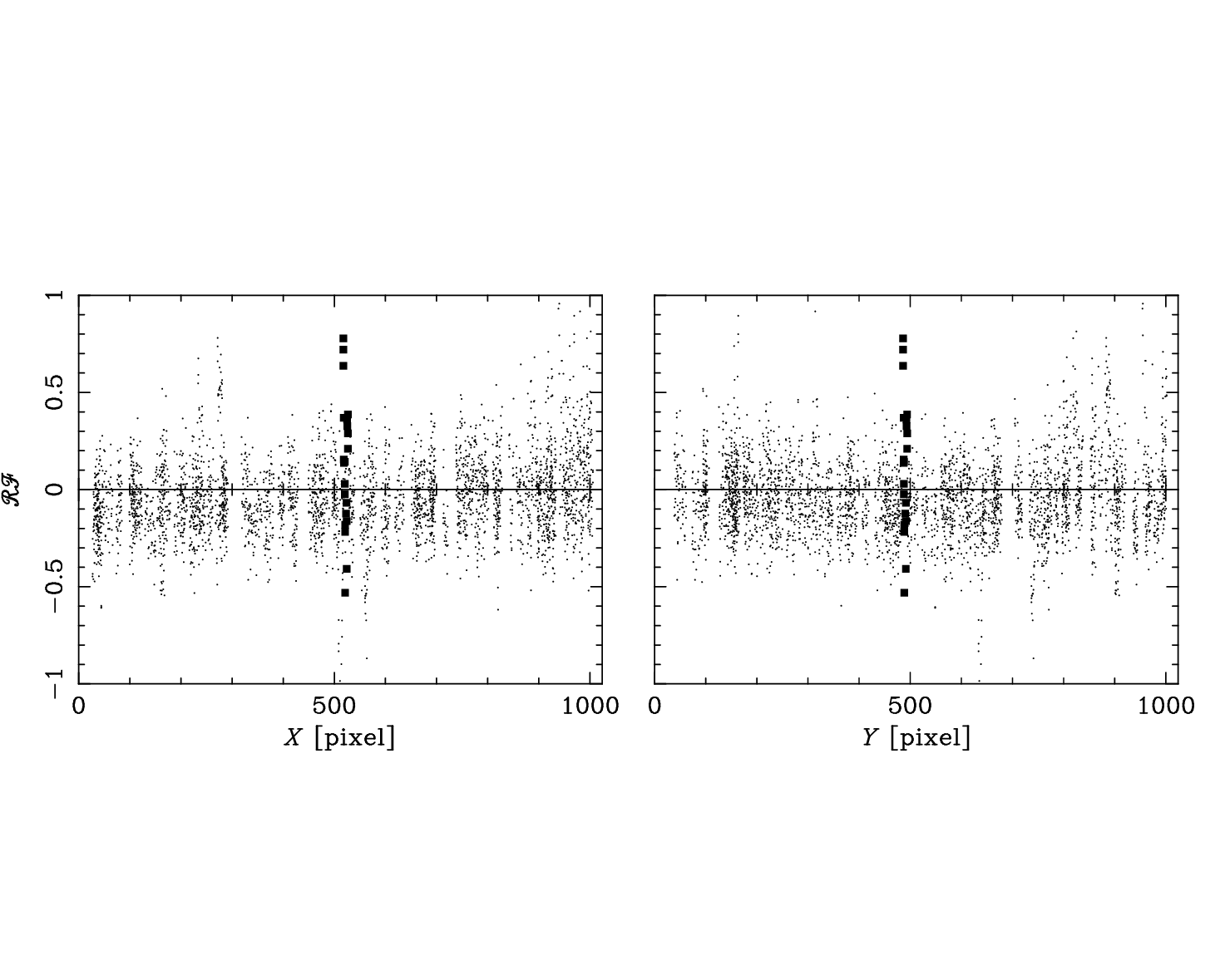}
}
\subfigure[]
{\label{stis-9-rf-x-2}
\includegraphics[angle=-90,scale=.32]{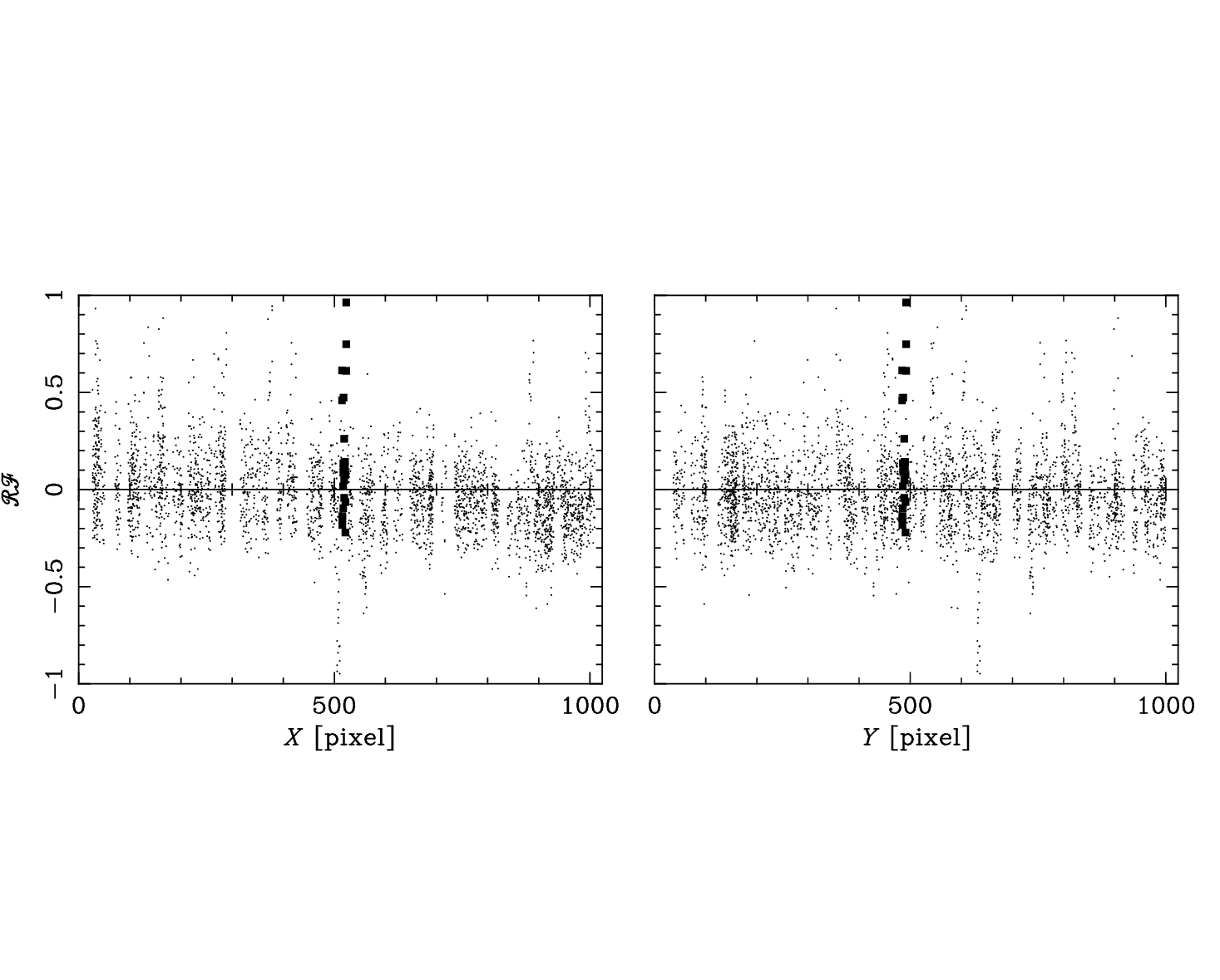}
}
\subfigure[]
{\label{stis-10-rf-x-2}
\includegraphics[angle=-90,scale=.32]{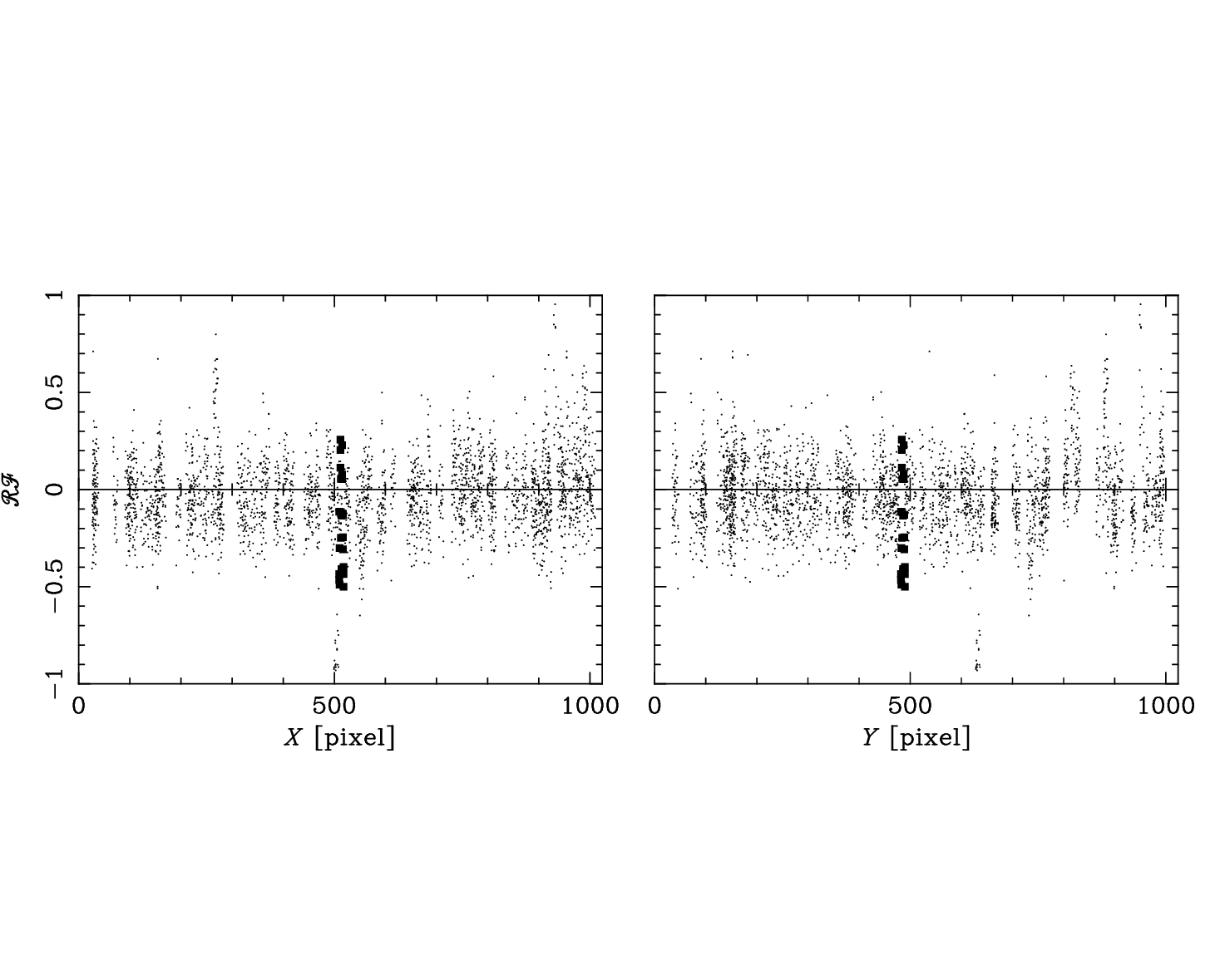}
}
\caption{Flux residual \textit{versus} location for objects in an image
for the SCL~$J0100-3338$ field.  Every plot displays points from 24
images.  The left-hand panels plot ${\cal RF}$ \textit{versus} the
$X$-coordinate and the right-hand panels plot ${\cal RF}$
\textit{versus} the $Y$-coordinate.  The filled squares represent the
QSO.  From top to bottom, the three rows of panels correspond to the
1999, 2000, and 2002 epoch, respectively.  For ease of comparison, all
of the plots have the same scale on the vertical axis.  Only objects
with a $S/N$ greater than 15 are plotted.}
\label{fig:rf-scl2}
\end{figure}

\clearpage
\begin{figure}
\centering
\subfigure[]
{\label{rx-ry-9-scl1}
\includegraphics[angle=-90,scale=.5]{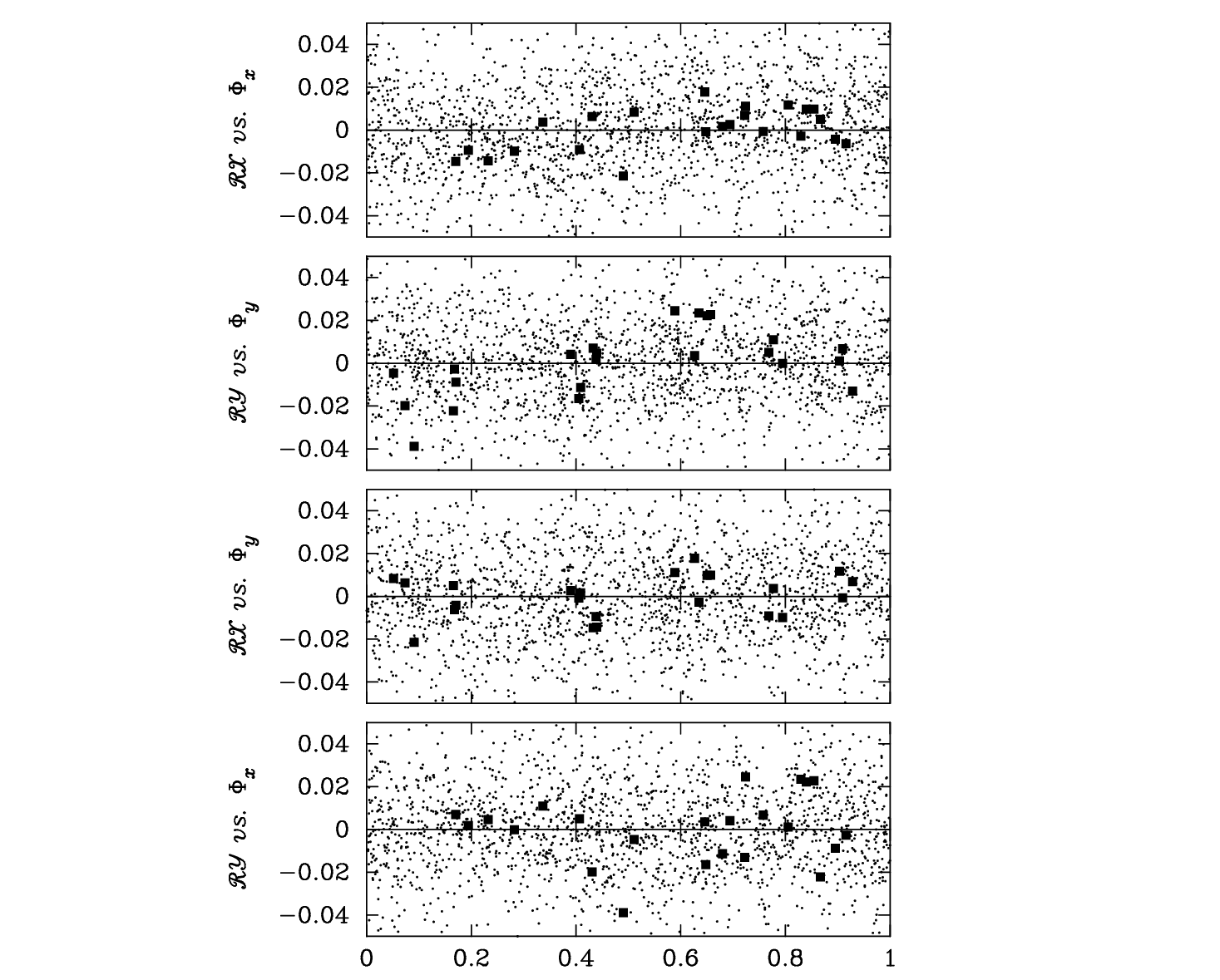}
}
\subfigure[]
{\label{rx-ry-10-scl1}
\includegraphics[angle=-90,scale=.5]{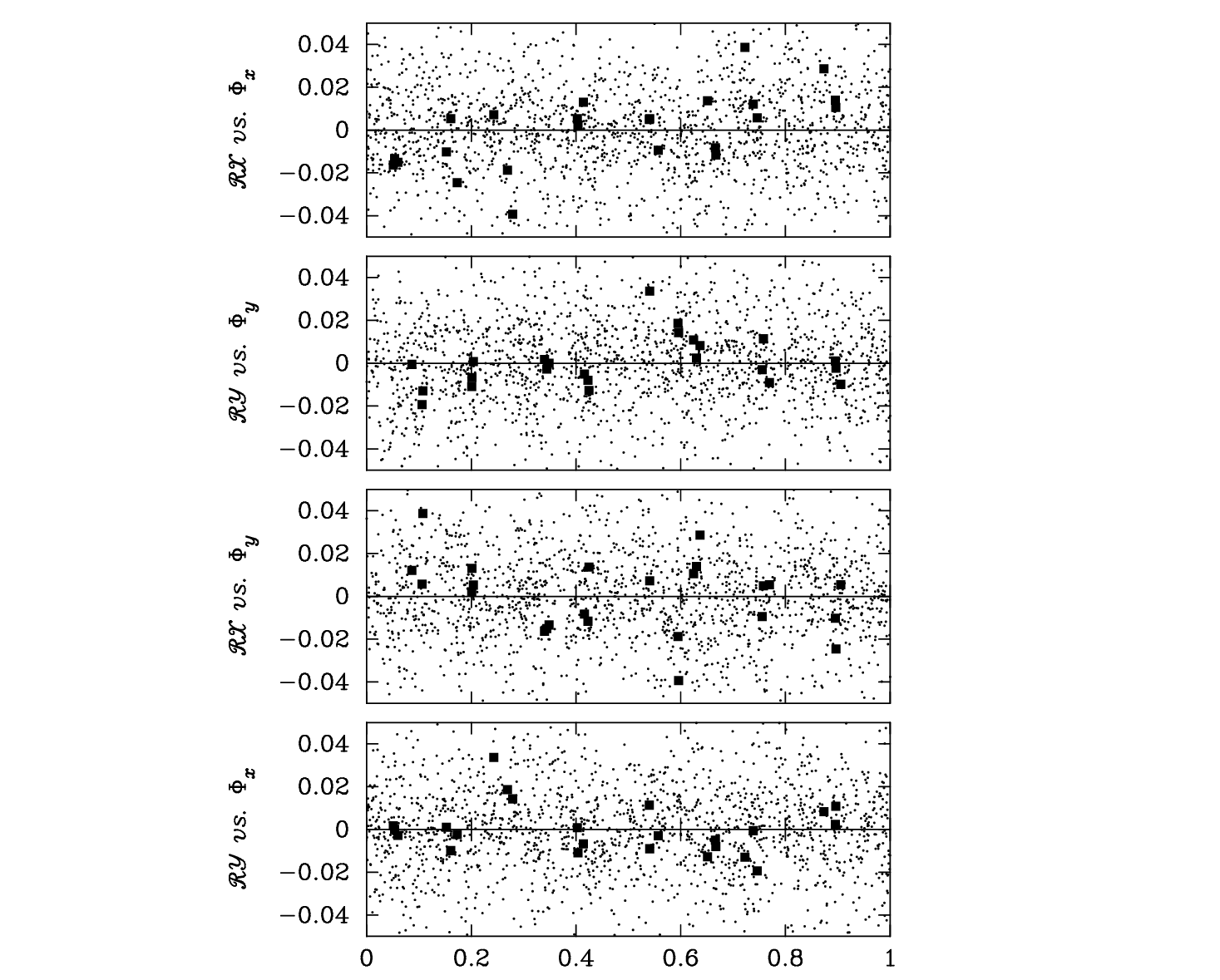}
}
\caption{Plots for the SCL~$J0100-3341$ field of the position
residuals, ${\cal RX}$ and ${\cal RY}$, as a function of the pixel
phase, $\Phi_{x}$ and $\Phi_{y}$.  The panels a) and b) correspond to
the epochs 2000 and 2002, respectively.  The filled squares
corresponds to the QSO.  Only objects with a $S/N$ greater than 30 are
plotted.}
\label{fig:rxry-scl1}
\end{figure}

\clearpage
\begin{figure}
\subfigure[]
{\label{rx-ry-8-scl2}
\includegraphics[angle=-90,scale=.4]{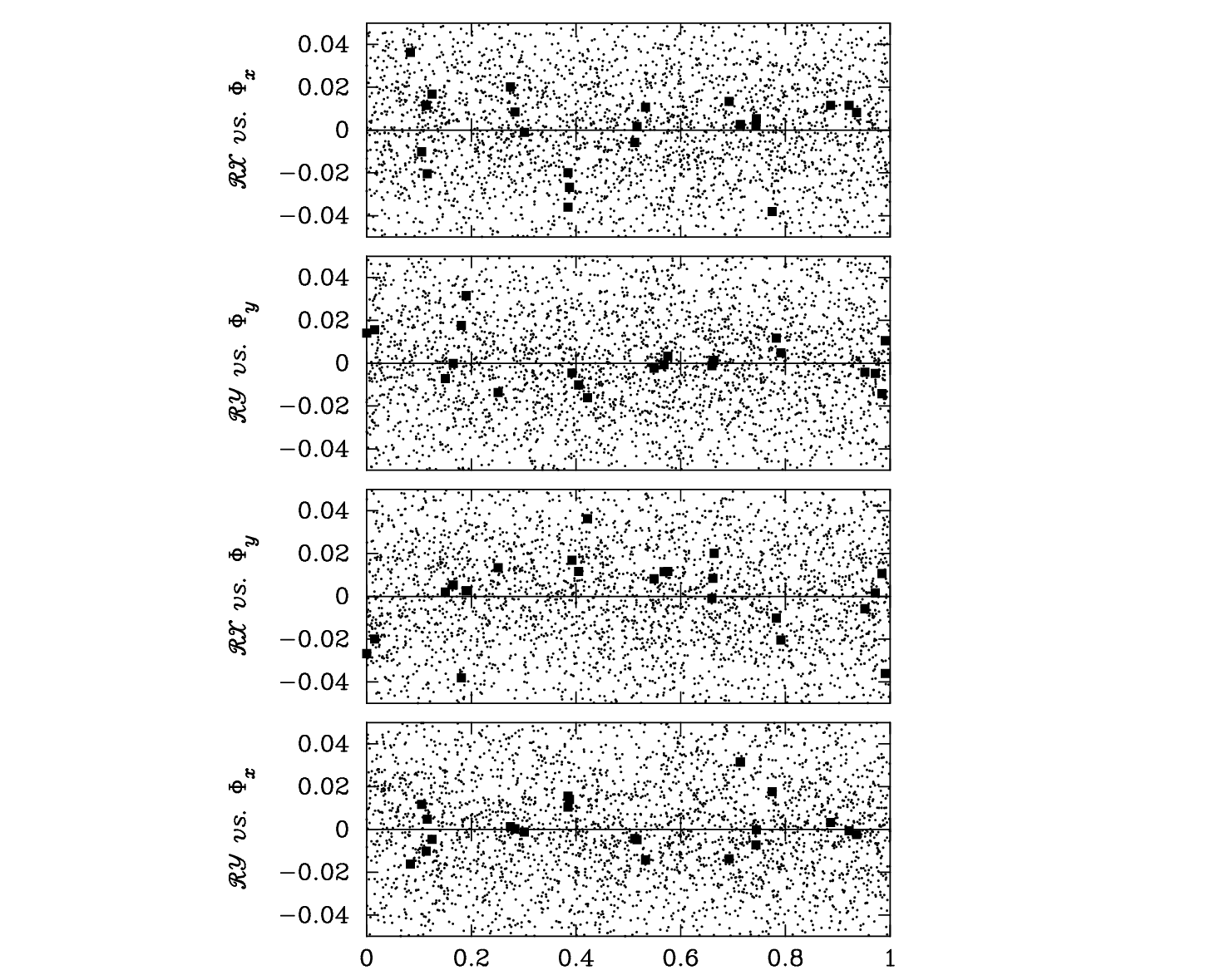}
}
\subfigure[]
{\label{rx-ry-9-scl2}
\includegraphics[angle=-90,scale=.4]{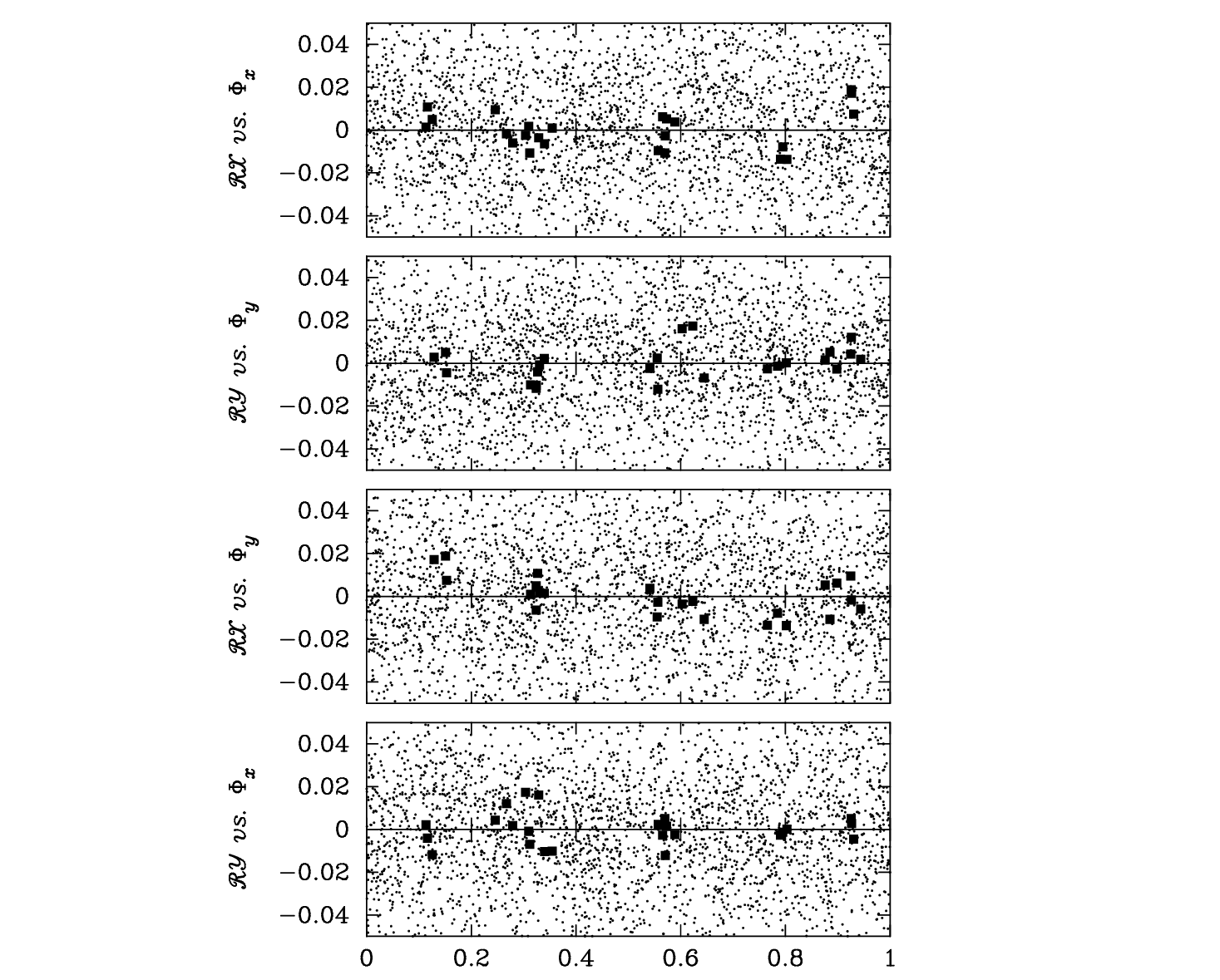}
}
\subfigure[]
{\label{rx-ry-10-scl2}
\includegraphics[angle=-90,scale=.4]{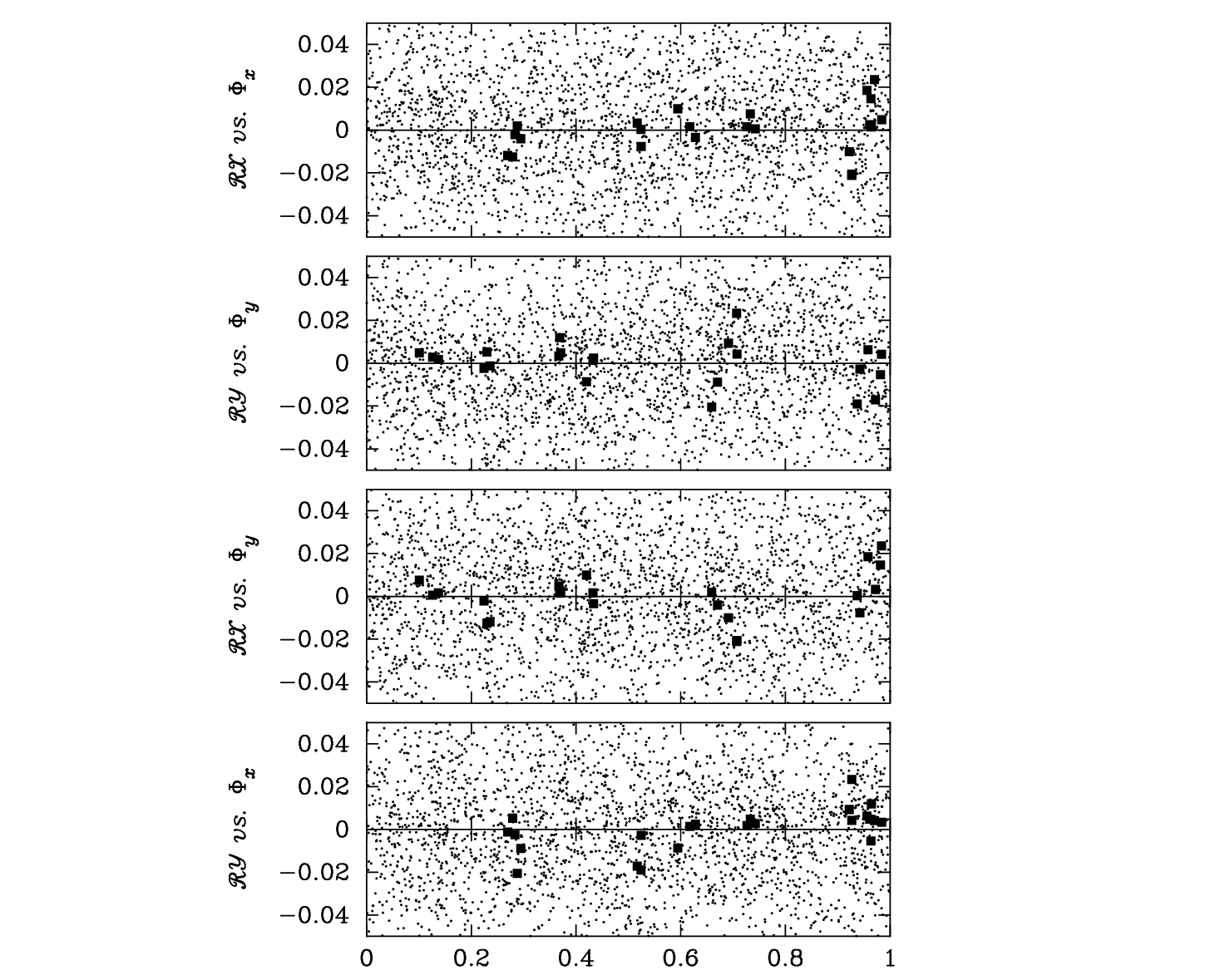}
}
\caption{Plots for the SCL~$J0100-3338$ field of the position
residuals, ${\cal RX}$ and ${\cal RY}$, as a function of the pixel
phase, $\Phi_{x}$ and $\Phi_{y}$.  The panels a), b), and c)
correspond to the epochs 1999, 2000, and 2002, respectively.  The
filled squares corresponds to the QSO.  Only objects with a $S/N$
greater than 15 are plotted.}
\label{fig:rxry-scl2}
\end{figure}

\clearpage
\begin{figure}
\includegraphics[angle=-90,scale=.8]{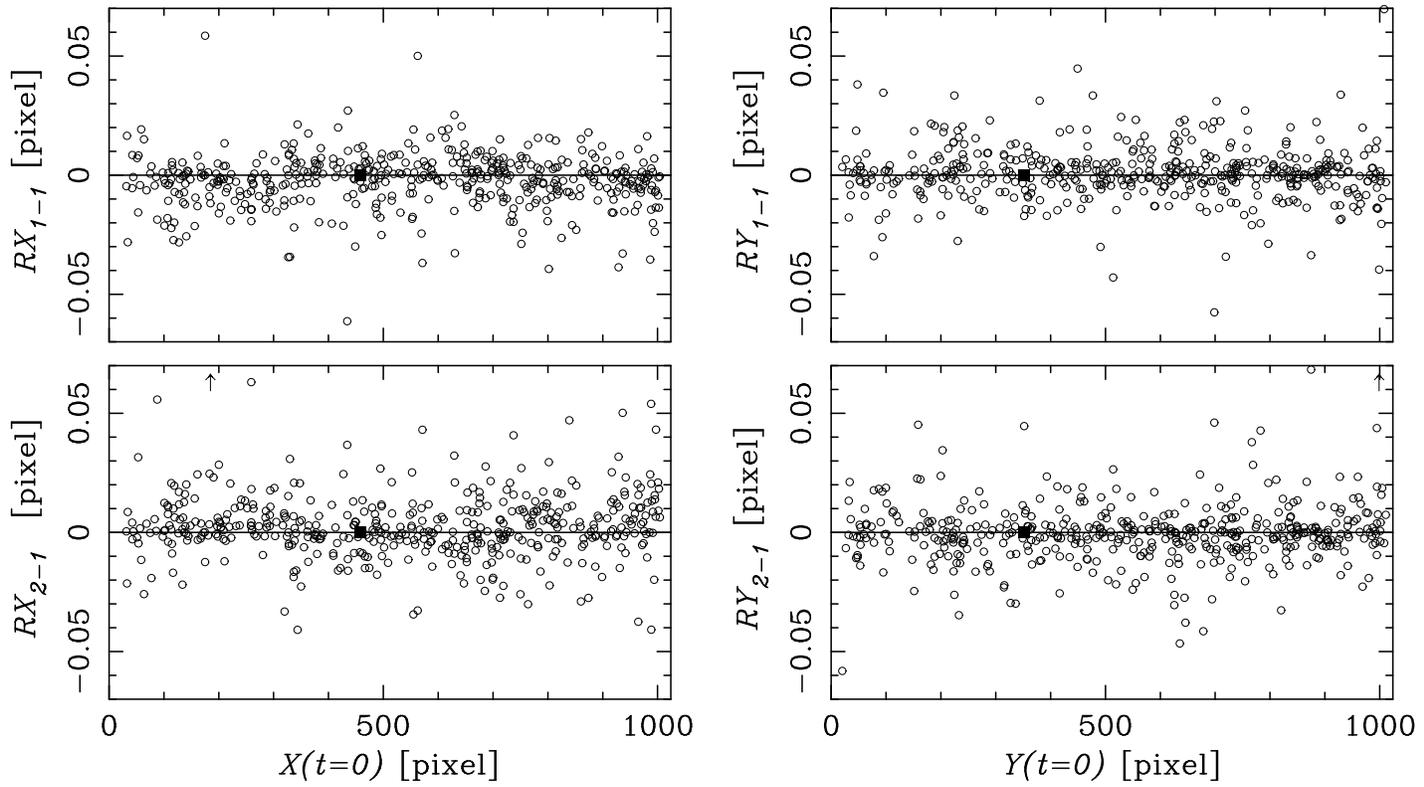}
\caption{Position residuals defined by the Equations~10 and 11 in P05
for the objects in the SCL~$J0100-3341$ field.  From top to bottom,
the panels are for the 2000 and 2002 epoch, respectively.  The panels
on the left show $RX$ \textit{versus} $X$ and those on the right show
$RY$ \textit{versus} $Y$.  The filled squares correspond to the QSO.
The arrows indicate points beyond the boundaries of the plot.  Note
the ``steps'' in the plots of $RX_{1-1}$ and $RX_{2-1}$
\textit{versus} $X$ at $X \simeq 320$~pixels.\label{fig:rx-ry-scl1}}
\end{figure}

\clearpage
\begin{figure}
\includegraphics[angle=-90,scale=.8]{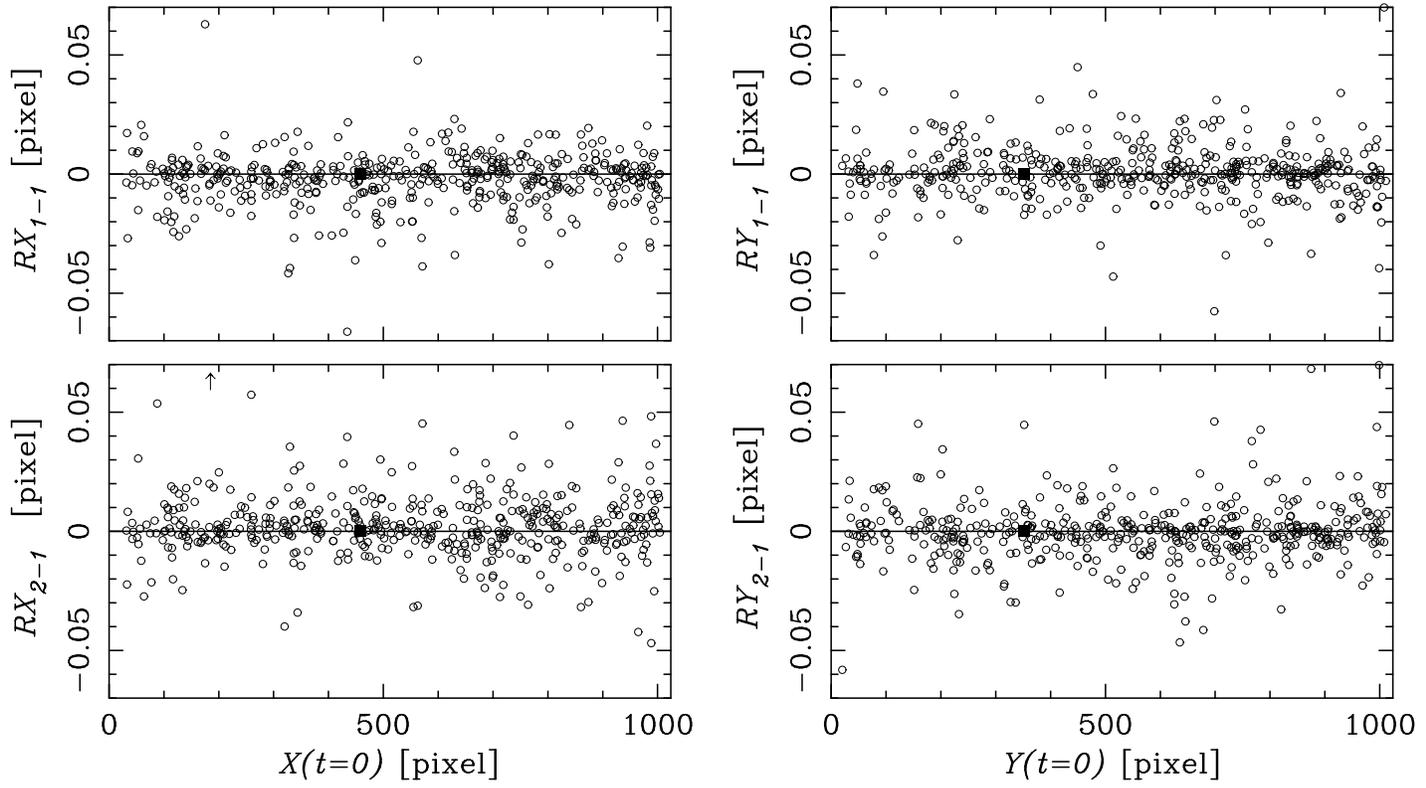}
\caption{The same as Figure~\ref{fig:rx-ry-scl1} after replacing
$x_{j}^{i}$ with $x_{j}^{i} + c_{7}$ in the transformation for $X \leq
320$~pixels.  The fitted value of $c_7$ is 0.019~pixel.  Note the
absence of ``steps.''\label{fig:rx-ry-scl1-c}}
\end{figure}

\clearpage
\begin{figure}
\includegraphics[angle=-90,scale=.8]{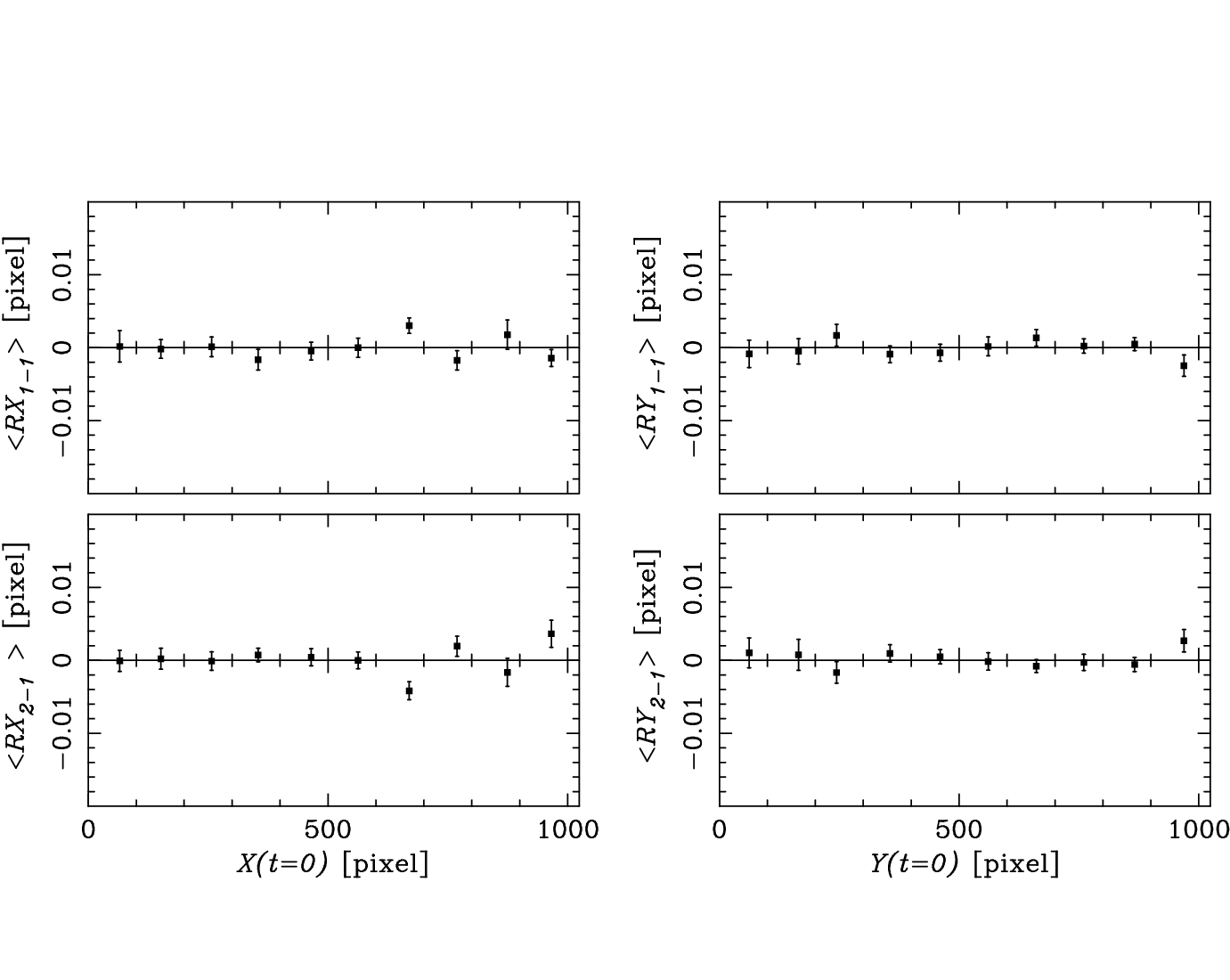}
\caption{The same as Figure~\ref{fig:rx-ry-scl1-c} except that the
points are the weighted mean residuals in ten equal-length bins in
$X$ or $Y$.  The points are plotted at the mean of the coordinate
values in the bin.
\label{fig:rx-ry-scl1-cb}}
\end{figure}

\clearpage
\begin{figure}
\includegraphics[angle=-90,scale=.8]{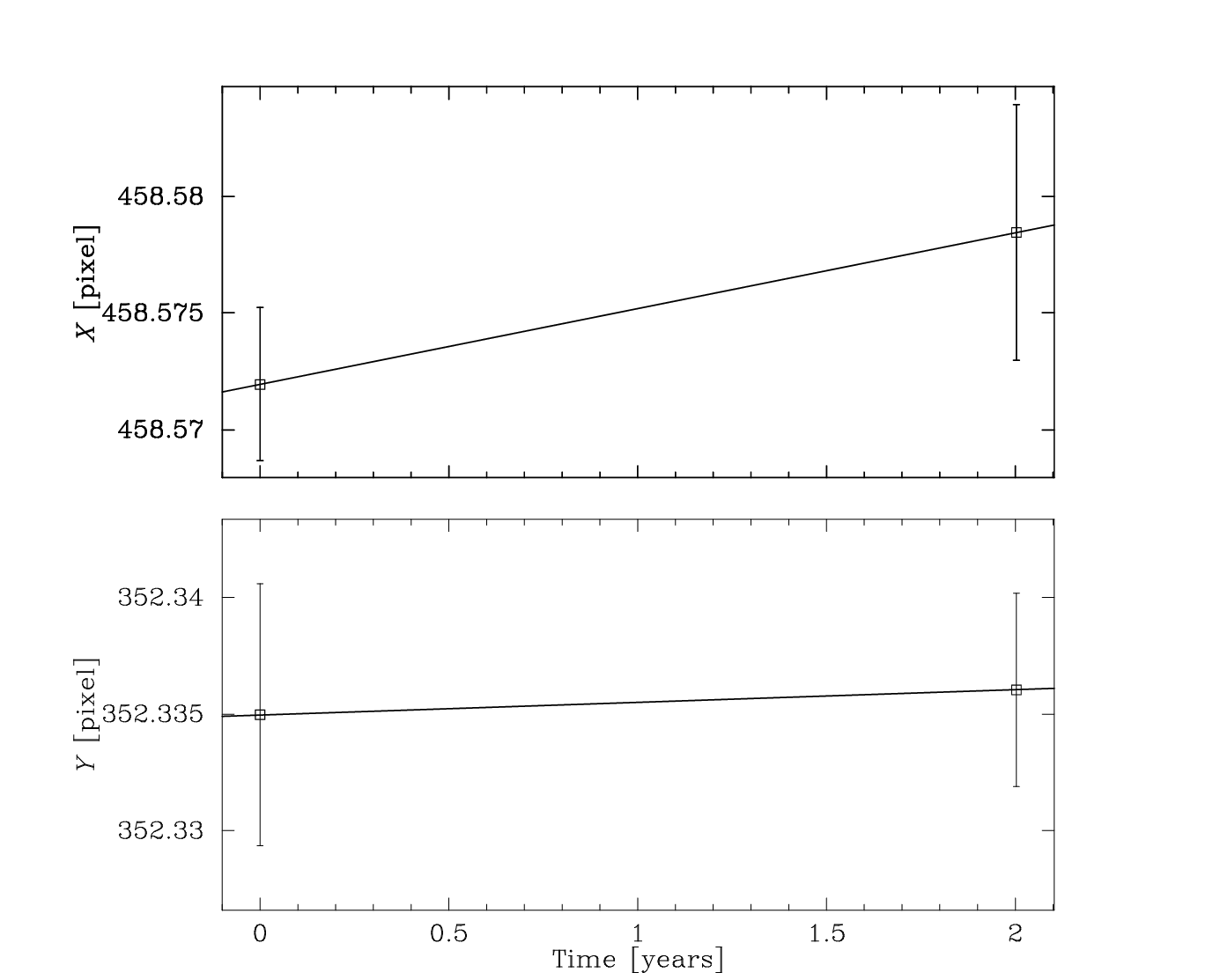}
\caption{The location of the QSO as a function of time for the
SCL~$J0100-3341$ field in the standard coordinate system.  The
vertical axis in each plot has the same scale.\label{fig:xq-yq-scl1}}
\end{figure}

\clearpage
\begin{figure}
\includegraphics[angle=-90,scale=.8]{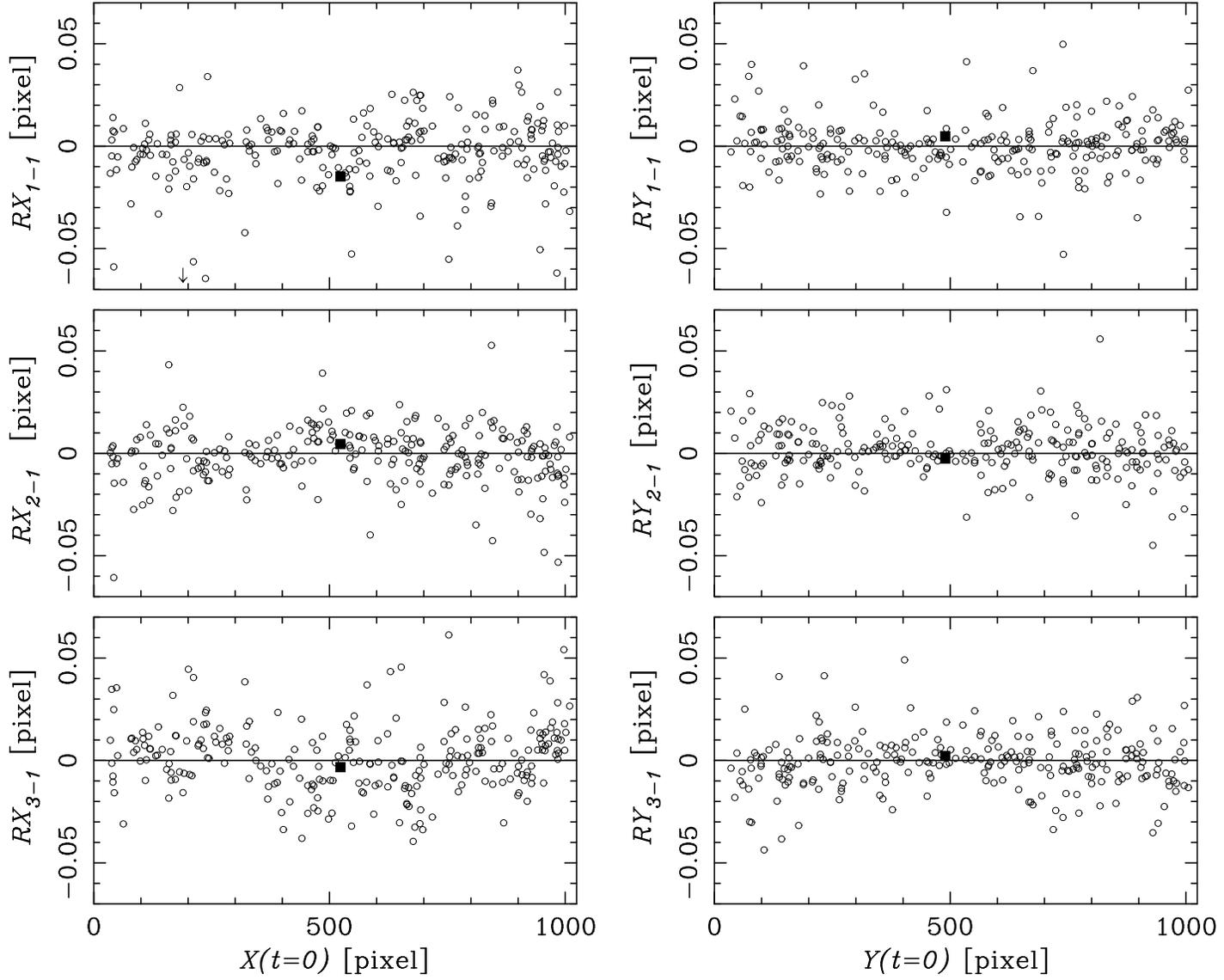}
\caption{Plots of the position residuals for the SCL~$J0100-3338$
field.  The panels in the left column show $RX$ \textit{versus} $X$
and those in the right column show $RY$ \textit{versus} $Y$. The
epochs increase chronologically from the top to the bottom row.  The
filled squares mark the QSO.  The arrow indicates a point beyond the
boundaries of the plot.\label{fig:rx-ry-scl2}}
\end{figure}

\clearpage
\begin{figure}
\includegraphics[angle=-90,scale=.8]{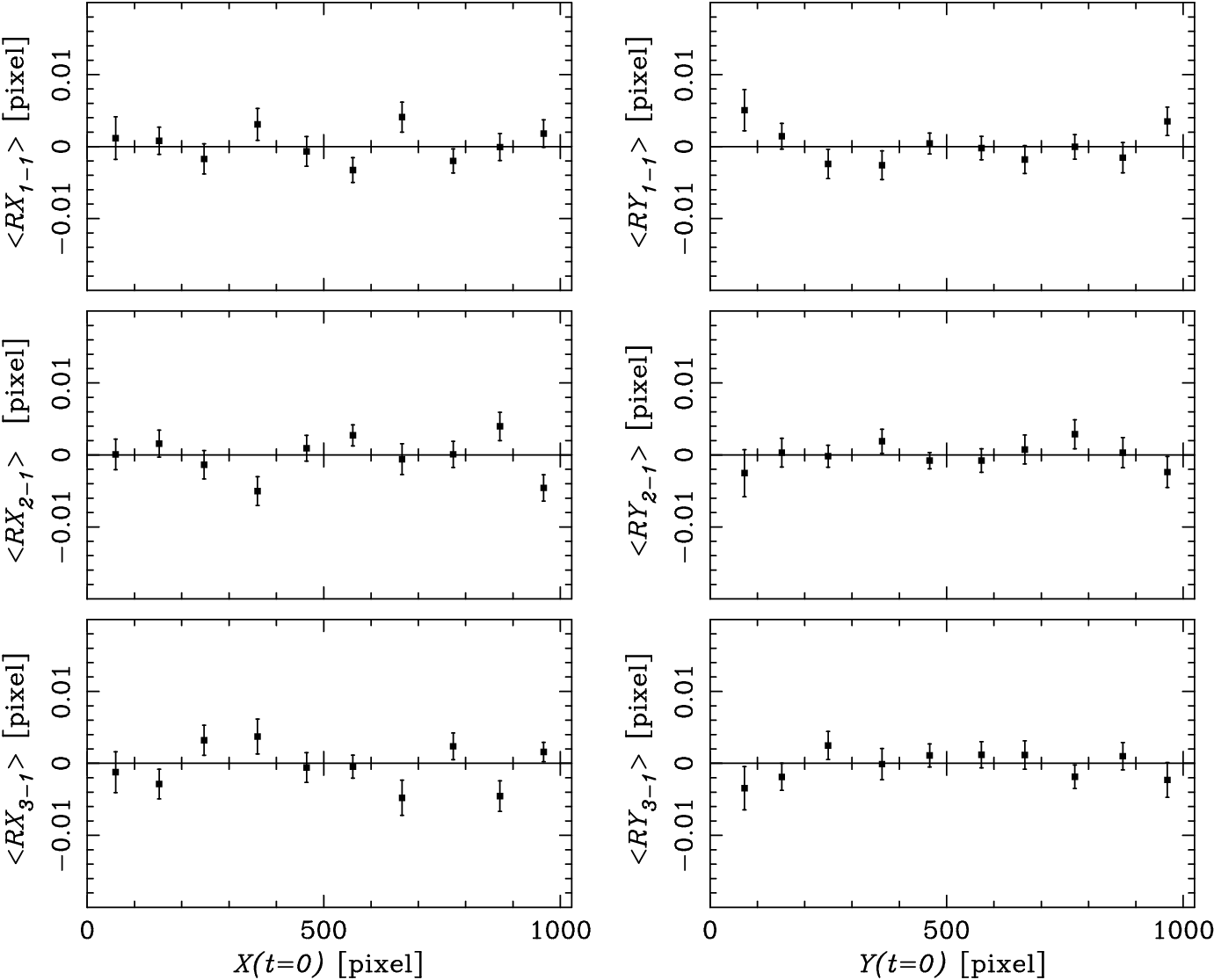}
\caption{The same as Figure~\ref{fig:rx-ry-scl2} except that the
points are the weighted mean residuals in ten equal-length bins in
$X$ or $Y$.  The points are plotted at the mean of the coordinate
values in the bin.
\label{fig:rx-ry-scl2-b}}
\end{figure}

\begin{figure}
\includegraphics[angle=-90,scale=.8]{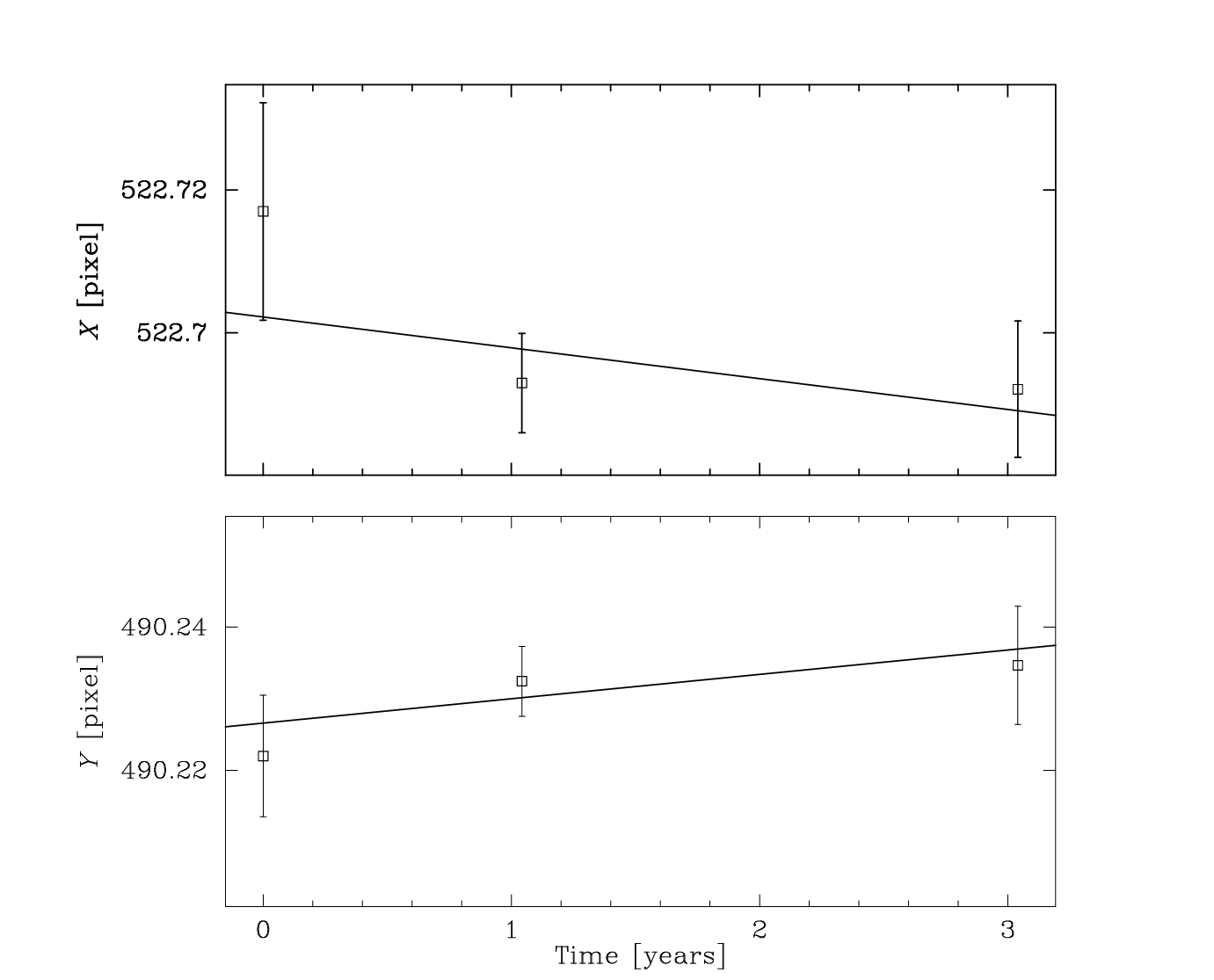}
\caption{The location of the QSO as a function of time for the
SCL~$J0100-3338$ field in the standard coordinate system.  The
vertical axis in each plot has the same scale.\label{fig:xq-yq-scl2}}
\end{figure}

\begin{figure}
\includegraphics[angle=-90,scale=.8]{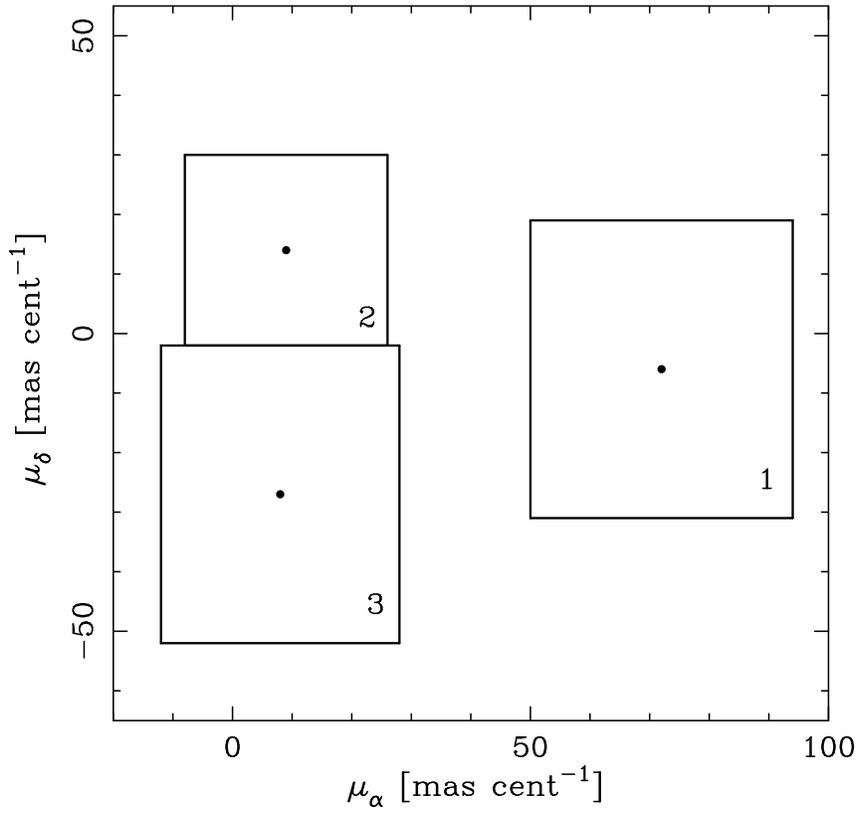}
\caption{Comparison of three independent measurements of the proper
motion for Sculptor.  The center of a rectangle --- marked with a dot
--- is the best estimate of the proper motion and the sides are offset
by the 1-$\sigma$ uncertainties.  Rectangles 1, 2, and 3 correspond to
the measurements from Schweitzer \etal\ (1995), this study (field
SCL~$J0100-3341$), and this study (field SCL~$J0100-3338$),
respectively.\label{fig:pm}}
\end{figure}

\begin{figure}
\includegraphics[angle=-90,scale=.8]{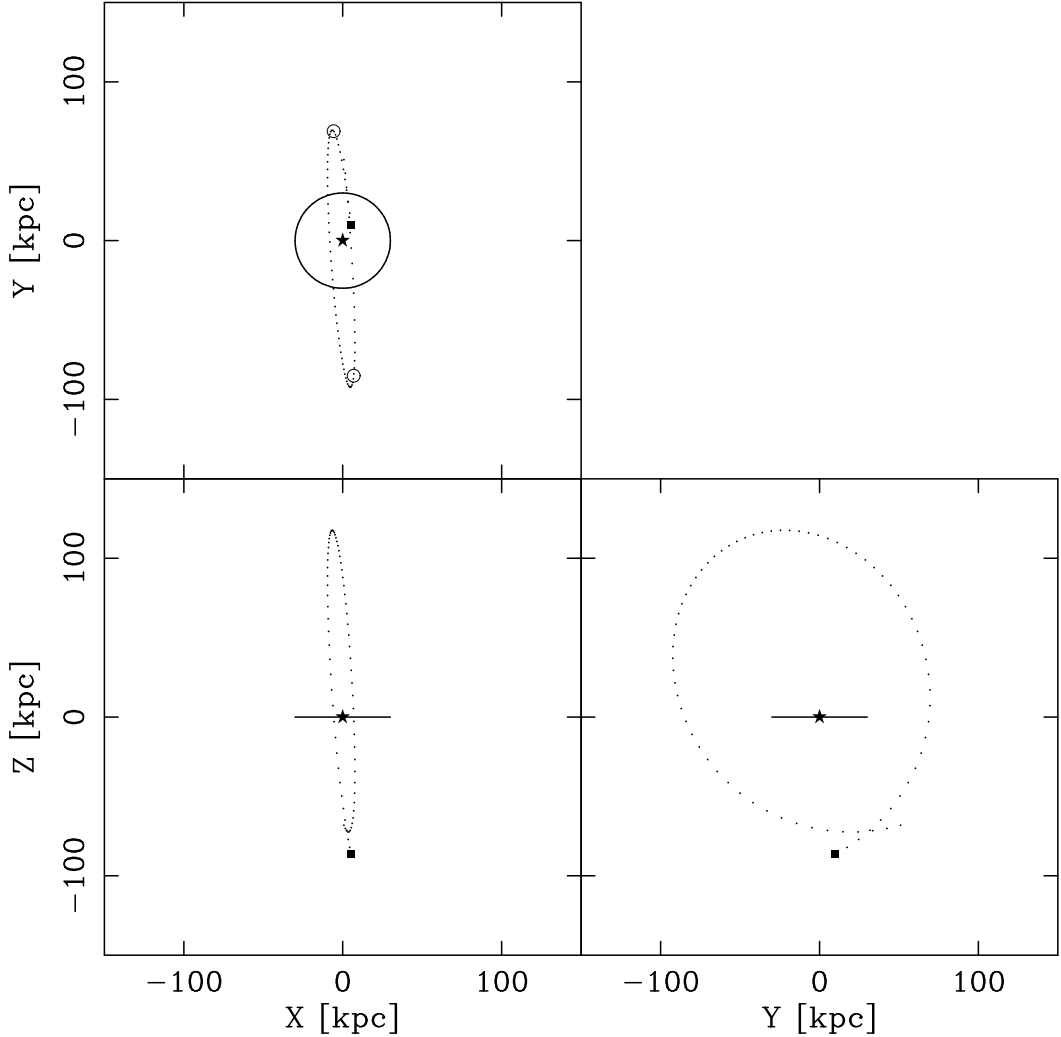}
\caption{Projections of the orbit of Sculptor onto the $X-Y$ plane
(top-left panel), the $X-Z$ plane (bottom-left panel), and the $Y-Z$
plane (bottom-right panel).  The origin of the right-handed coordinate
system is at the Galactic center, which is marked with a filled star.
The Galactic disk is in the $X-Y$ plane and the present location of
the Sun is on the positive $X$ axis.  The filled square marks the
current location of Sculptor at $(X,Y,Z)=(5.4,9.9,-86)$~kpc.  For
reference, the large circle in the $X-Y$ plane has a radius of 30~kpc.
The three small circles in the $X-Y$ projection mark the points where
Sculptor passes through the plane of the Galactic disk.  The
integration starts from the present and extends backwards in time for
3~Gyr.\label{fig:orbit}}
\end{figure}

\clearpage
\begin{figure}
\includegraphics[angle=-90,scale=.8]{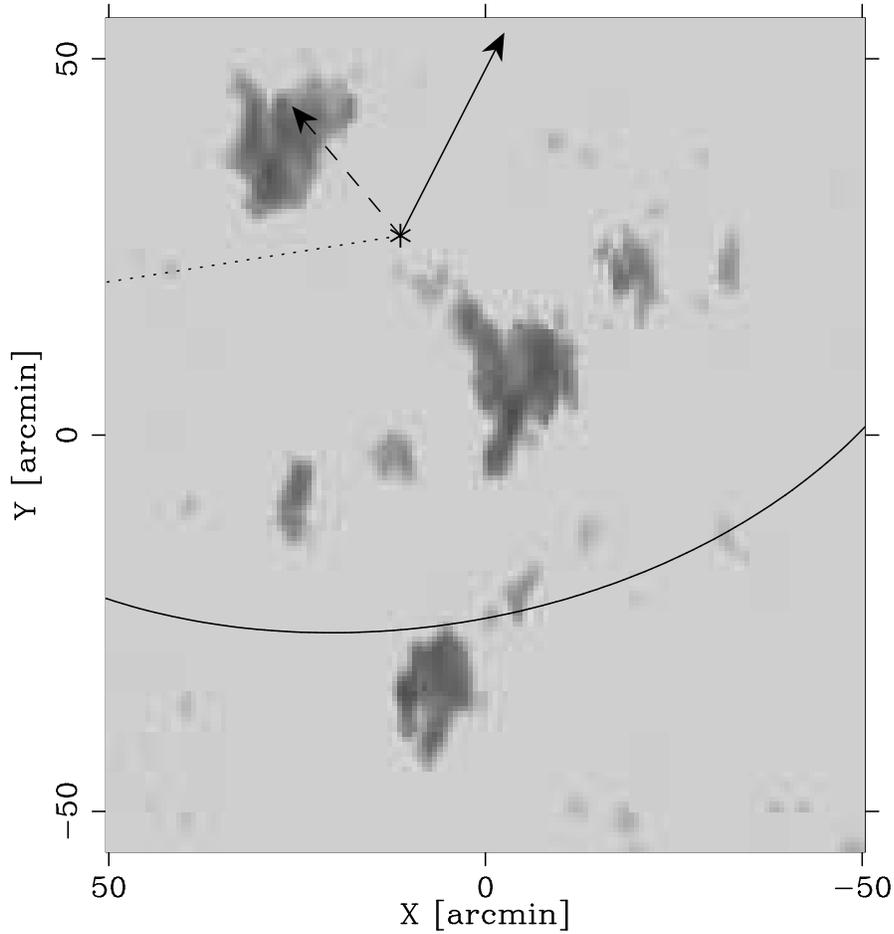}
\caption{\textit{Gray-scale map:} The distribution of HI in the
direction of Sculptor based on data taken with the Australia Telescope
Compact Array by Bouchard \etal\ (2003).  The map is centered at
$(X,Y)=(0,0)$, which corresponds to $(\alpha, \delta) =
(00^{\mbox{h}}59^{\mbox{m}}18^{\mbox{s}}, -34^{\circ}
10^{\prime}00^{\prime\prime})$ (J2000.0).  North is up and East is to
the left.  Overplotted on the map are: the center of Sculptor
(asterisk); its optical boundary (ellipse); the galactic rest frame
proper motion derived by Schweitzer \etal\ (1995) (dashed arrow) and
by us (solid arrow); and the line connecting the centers of Sculptor
and Fornax (dotted line).\label{fig:gas}}
\end{figure}

\clearpage

\setcounter{table}{0}
\newdimen\digitwidth\setbox0=\hbox{\rm 0}\digitwidth=\wd0
\catcode`@=\active\def@{\kern\digitwidth}
\begin{deluxetable}{lrr}
\tablecolumns{3}
\tablewidth{3.5truein}
\tablecaption{Measured Proper Motion of Sculptor}
\tablehead{
&\colhead{$\mu_{\alpha}$}&\colhead{$\mu_{\delta}$}\\ 
\colhead{Field}&\multicolumn{2}{c}{(mas cent$^{-1}$)}\\
\colhead{(1)}&\colhead{(2)}&\colhead{(3)}}
\startdata
SCL@J$0100-3341$ &$9\pm17$&$14\pm16$ \\
\noalign{\vspace{1pt}}
SCL@J$0100-3338$ &$8\pm20$&$-27\pm25$ \\
\hline
\noalign{\vspace{1pt}}
Weighted mean&$9\pm13$&$2\pm13$\\
\enddata
\end{deluxetable}

\begin{deluxetable}{ccccccc}
\tablecolumns{7}
\tablewidth{5.0truein} 
\tablecaption{Measured Proper Motions For Objects in
the SCL~$J0100-3341$ Field}
\tablehead{ &X&Y& &$\mu_{\alpha}$&$\mu_{\delta}$ &  \\
\colhead{ID}&\colhead{(pixels)}&\colhead{(pixels)}&\colhead{$S/N$}&\colhead{(mas
@cent$^{-1}$)}&
\colhead{(mas@cent$^{-1}$)} & \colhead{$\chi^2$} \\
\colhead{(1)}&\colhead{(2)}&\colhead{(3)}&\colhead{(4)}&\colhead{(5)}&
\colhead{(6)} & \colhead{(7)}
}
\startdata
  1& 459& 352& 125& $    0 \pm  21 $& $    0 \pm  21 $&  \nodata \\
  2& 327& 740& 164& $ 2766 \pm  17 $& $ 1048 \pm  18 $&  \nodata \\
  3& 588& 330& 112& $ -812 \pm  16 $& $-1714 \pm  17 $&  \nodata \\
  4& 137& 627&  99& $  180 \pm  17 $& $ -127 \pm  21 $&  \nodata \\
  5& 496& 398&  10& $  -66 \pm  46 $& $  178 \pm  72 $&  \nodata \\
  6& 574& 710&   6& $   36 \pm  93 $& $  258 \pm  63 $&  \nodata \\
\enddata
\end{deluxetable}

\begin{deluxetable}{ccccccc}
\tablecolumns{7}
\tablewidth{5.0truein} 
\tablecaption{Measured Proper Motions For Objects in
the SCL~$J0100-3338$ Field}
\tablehead{ &X&Y& &$\mu_{\alpha}$&$\mu_{\delta}$ &  \\
\colhead{ID}&\colhead{(pixels)}&\colhead{(pixels)}&\colhead{$S/N$}&\colhead{(mas
@cent$^{-1}$)}&
\colhead{(mas@cent$^{-1}$)} & \colhead{$\chi^2$} \\
\colhead{(1)}&\colhead{(2)}&\colhead{(3)}&\colhead{(4)}&\colhead{(5)}&
\colhead{(6)} & \colhead{(7)}
}
\startdata
  1& 523& 490& 180& $    0 \pm  24 $& $    0 \pm  28 $&  2.10 \\
  2& 414& 177&  42& $ -188 \pm  22 $& $ -299 \pm  20 $&  0.61 \\
  3& 950& 832&  20& $   89 \pm  31 $& $ -562 \pm  27 $&  4.25 \\
\enddata
\end{deluxetable}

\hoffset=-0.5truein
\begin{deluxetable}{lrrrrrrrrr}
\tablecolumns{10}
\tablewidth{8.8truein} 
\tablecaption{Galactic-Rest-Frame Proper Motion and Space Velocity of Sculptor}
\tablehead{&\colhead{$\mu_{\alpha}^{\mbox{\tiny{Grf}}}$}&
\colhead{$\mu_{\delta}^{\mbox{\tiny{Grf}}}$}&
\colhead{$\mu_{l}^{\mbox{\tiny{Grf}}}$}&\colhead{$\mu_{b}^{\mbox{\tiny{Grf}}}$}&
\colhead{$\Pi$}&\colhead{$\Theta$}&\colhead{$Z$}&\colhead{$V_{r}$}&
\colhead{$V_{t}$}\\ 
\colhead{Field}&\multicolumn{2}{c}{(mas cent$^{-1}$)}
&\multicolumn{2}{c}{(mas cent$^{-1}$)}&\colhead{(km s$^{-1}$)}&\colhead{(km s$^{-1}$)}&\colhead{(km s$^{-1}$)}
&\colhead{(km s$^{-1}$)}&\colhead{(km s$^{-1}$)}\\
\colhead{(1)}&\colhead{(2)}&\colhead{(3)}&\colhead{(4)}
&\colhead{(5)}&\colhead{(6)}&\colhead{(7)}&\colhead{(8)}&\colhead{(9)}
&\colhead{(10)}}
\startdata
SCL@J$0100-3341$&$-23\pm17$&$57\pm16$& $6\pm17$&$-62\pm17$& $-186\pm69$&$159\pm68$&$-107\pm8$& $82\pm7$&$254\pm66$\\
\noalign{\vspace{5pt}}
SCL@J$0100-3338$&$-24\pm20$&$17\pm25$& $18\pm21$&$-23\pm25$& $-113\pm88$&$10\pm100$&$-88\pm12$& $73\pm9$&$124\pm76$\\
\noalign{\vspace{3pt}}
\hline
\noalign{\vspace{5pt}}
Weighted mean&$-23\pm13$&$45\pm13$& $11\pm13$&$-50\pm14$& $-158\pm54$&$112\pm56$&$-101\pm7$& $79\pm6$&$198\pm50$\\
\enddata
\end{deluxetable}

\begin{deluxetable}{lcccc}
\tablecolumns{5}
\tablewidth{4.75truein} 
\tablecaption{Orbital elements of Sculptor}
\tablehead{Quantity&Symbol&Unit&Value&95\% Conf. Interv.\\
\colhead{(1)}&\colhead{(2)}&\colhead{(3)}&\colhead{(4)}&\colhead{(5)}}
\startdata
Perigalacticon&$R_{p}$&kpc&$68$&$(31,83)$ \\
\noalign{\vspace{1pt}}
Apogalacticon&$R_{a}$&kpc&$122$&$(97,313)$\\
\noalign{\vspace{1pt}}
Eccentricity &$e$&&$0.29$&$(0.26,0.60)$\\
\noalign{\vspace{1pt}}
Period&$T$&Gyr&$2.2$&$(1.5,4.9)$\\
\noalign{\vspace{1pt}}
Inclination&$\Phi$&deg&86&$(83,90)$ \\
\noalign{\vspace{1pt}}
Longitude&$\Omega$&deg&275&$(243,306)$ \\
\enddata
\end{deluxetable}

\begin{deluxetable}{lcccc}
\tablecolumns{5}
\tablewidth{4.5truein} 
\tablecaption{Predicted Proper Motion of Sculptor}
\tablehead{
&
\colhead{$\mu_{\alpha}$}&
\colhead{$\mu_{\delta}$}&
\colhead{$\vert \vec{\mu}\vert$}&
\colhead{PA}\\
\colhead{Stream No.}&\multicolumn{3}{c}{(mas
cent$^{-1}$)}&\colhead{(degrees)}\\ 
\colhead{(1)}&\colhead{(2)}&\colhead{(3)}&\colhead{(4)}&\colhead{(5)}}
\startdata
2  & 51  & --80 & 95 & 147 \\
\noalign{\vspace{1pt}}
4a & 80  & --61  & 101 & 127 \\
\noalign{\vspace{1pt}}
4b & -13 & -27 & 30 & 205\\
\noalign{\vspace{1pt}}
\hline
\noalign{\vspace{1pt}}
Our Result&$9\pm13$&$2\pm13$&$9\pm13$&$77\pm81$ \\
\enddata
\end{deluxetable}
\end{document}